\documentclass[prd,twocolumn,showpacs,showkeys,preprintnumbers,floatfix,
  nofootinbib,superscriptaddress]{revtex4-1}
\usepackage{amsfonts} 
\usepackage{amssymb} 
\usepackage{amsmath} 
\usepackage{graphicx} 
\usepackage[FIGBOTCAP,TABBOTCAP,tight]{subfigure}
\usepackage{array} 
\usepackage{dcolumn} 
\usepackage{bm} 
\usepackage{latexsym} 
\usepackage{longtable} 
\usepackage{hyperref} 
\usepackage{multirow}
\usepackage{xcolor}
\usepackage{verbatim}

\newcommand{\MeV}{\mathop{\rm MeV}\nolimits}
\newcommand{\GeV}{\mathop{\rm GeV}\nolimits}

\newcommand{\MSbar}{\overline{\mathrm{MS}}}

\newcommand{\ben}[1]{#1} 

\newcommand{\wlee}[1]{#1} 

\newcommand{\sharpe}[1]{#1} 

\newcommand{\com}[1]{} 

\allowdisplaybreaks

\begin{document}

\title{Kaon BSM B-parameters using 
  improved staggered fermions from $N_f=2+1$ unquenched QCD }
\author{\ben{Benjamin J. Choi}}
\affiliation{
	Lattice Gauge Theory Research Center, FPRD, and CTP, \\
	Department of Physics and Astronomy,
	Seoul National University, Seoul 08826, South Korea
}
\author{Yong-Chull Jang}
\affiliation{
  Lattice Gauge Theory Research Center, FPRD, and CTP, \\
  Department of Physics and Astronomy,
  Seoul National University, Seoul 08826, South Korea
}
\author{Chulwoo Jung}
\email[E-mail: ]{chulwoo@bnl.gov}
%
%
\affiliation{
  Physics Department, Brookhaven National Laboratory,
  Upton, NY11973, USA
}
\author{Hwancheol Jeong}
\affiliation{
  Lattice Gauge Theory Research Center, FPRD, and CTP, \\
  Department of Physics and Astronomy,
  Seoul National University, Seoul 08826, South Korea
}
\author{Jangho Kim}
\affiliation{
  National Institute of Supercomputing and Networking,\\ 
  Korea Institute of Science and Technology Information, 
  Daejeon 34141, South Korea
}
\author{Jongjeong Kim}
\affiliation{
  Lattice Gauge Theory Research Center, FPRD, and CTP, \\
  Department of Physics and Astronomy,
  Seoul National University, Seoul 08826, South Korea
}
\author{Sunghee Kim}
\affiliation{
  Lattice Gauge Theory Research Center, FPRD, and CTP, \\
  Department of Physics and Astronomy,
  Seoul National University, Seoul 08826, South Korea
}
\author{Weonjong Lee}
\email[E-mail: ]{wlee@snu.ac.kr}
%
%
%
\affiliation{
  Lattice Gauge Theory Research Center, FPRD, and CTP, \\
  Department of Physics and Astronomy,
  Seoul National University, Seoul 08826, South Korea
}
\author{Jaehoon Leem}
\affiliation{
  Lattice Gauge Theory Research Center, FPRD, and CTP, \\
  Department of Physics and Astronomy,
  Seoul National University, Seoul 08826, South Korea
}
\author{Jeonghwan Pak}
\affiliation{
  Lattice Gauge Theory Research Center, FPRD, and CTP, \\
  Department of Physics and Astronomy,
  Seoul National University, Seoul 08826, South Korea
}
\author{Sungwoo Park}
\affiliation{
  Lattice Gauge Theory Research Center, FPRD, and CTP, \\
  Department of Physics and Astronomy,
  Seoul National University, Seoul 08826, South Korea
}
\author{Stephen R. Sharpe}
\email[E-mail: ]{srsharpe@uw.edu}
\affiliation{
  Physics Department,
  University of Washington,
  Seattle, WA 98195-1560, USA
}
\author{Boram Yoon}
\affiliation{
   Los Alamos National Laboratory,
   Theoretical Division T-2, MS B283,
   Los Alamos, NM 87545, USA
 }
\collaboration{SWME Collaboration}
\date{\today}
\begin{abstract}
  We present results for the matrix elements of the additional $\Delta S=2$
  operators that appear in models of physics beyond the Standard Model (BSM),
  expressed in terms of four BSM $B$-parameters.
  Combined with experimental results for $\Delta M_K$ and $\epsilon_K$,
  these constrain the parameters of BSM models.
  We use improved staggered fermions, with valence HYP-smeared quarks and
  $N_f=2+1$ flavors of ``asqtad" sea quarks. The configurations have been generated
  by the MILC collaboration.
  The matching between lattice and continuum four-fermion operators and bilinears
  is done perturbatively at one-loop order.
%
%
  We use three lattice spacings for the continuum extrapolation: $a\approx 0.09$, $0.06$ and
  $0.045\;$fm. Valence light-quark masses range down to $\approx m_s^{\rm phys}/13$
  while the light sea-quark masses range down to $\approx m_s^{\rm phys}/20$.
  Compared to our previous published work, we have added four additional lattice
  ensembles, leading to better controlled extrapolations in the lattice spacing and
  sea-quark masses.
  We report final results for two renormalization scales, $\mu=2\;\text{GeV}$ and $3\;\text{GeV}$,
  and compare them to those obtained by other collaborations.
  Agreement is found for two of the four BSM $B$-parameters ($B_2$ and $B_3^\text{SUSY}$).
The other two ($B_4$ and $B_5$) differ significantly
  from those obtained using RI-MOM renormalization as an intermediate scheme,
  but are in agreement with recent preliminary results obtained by
  the RBC-UKQCD collaboration using RI-SMOM intermediate schemes.
\end{abstract}
\maketitle

\section{Introduction}
\label{sec:intr}

Neutral kaon mixing and the associated indirect
CP-violation have long provided an important window
into physics at high energy scales. 
In the Standard Model (SM), for example, 
the measured CP-violating parameter $\epsilon_K$
is sensitive to scales up to the top-quark mass.
To determine whether the measured value is consistent with
the SM, however, requires knowledge of the hadronic matrix
element parametrized by the kaon $B$-parameter, $B_K$.
Recently, lattice QCD calculations have matured to the point
that such matrix elements can be determined from first principles
with percent-level accuracy.\footnote{%
For a recent review of such quantities and their associated errors,
see Ref.~\cite{FLAG2}.}
Specifically, results for $B_K$ from Refs.~\cite{Blum:2014tka, 
Carrasco:2015pra, Durr:2011ap, Aubin2010:PRD,
Laiho:2011np, Bae:2010ki, Bae:2011ff, Bae:2014sja}
are such that the average has an error of $\sim 1.3\%$~\cite{FLAG2}.
This is accurate enough to provide strong constraints on SM parameters
(see, e.g., Refs.~\cite{Bailey:2015tba,Laiho:2009eu}).
Ultimately, lattice calculations will also be able to use the $K_L-K_S$ mass
difference, $\Delta M_K$, to test the SM~\cite{Bai:2014cva}.

Physics beyond the SM (BSM) will, in general, contribute to flavor changing
neutral processes such as kaon mixing. Indeed, unless there is some
cancellation akin to the GIM mechanism, rough estimates show that the
scale of new physics must be $\gtrsim 10^5\;$TeV
in order to avoid overly large contributions to $\Delta M_K$ and 
$\epsilon_K$~\cite{Bertone:2012cu}.
In fact, many BSM models have partial cancellations such that the
scale of new physics is accessible at the Large Hadron Collider (LHC),
but often such models are pushing against the constraints from kaon mixing.
If evidence for new physics is discovered at the LHC in the coming years, 
then, in order to sift through the available models,
it will be essential to turn the constraints from kaon mixing into
precision tools. To do this it is necessary to calculate the hadronic matrix
elements of the full basis of $\Delta S=2$ four-fermion operators that can appear.
Illustrations of how these matrix elements constrain BSM models are
given in Refs.~\cite{Ciuchini:1998ix, Bona:2005eu, Isidori:2010kg, Blanke:2011ry}.

In the SM, four-fermion operators in the effective $\Delta S=2$ Hamiltonian
are composed of left-handed currents.
Generic BSM physics, by contrast, also includes heavy virtual particles
coupling to right-handed quarks.
Because of this, the single ``left-left" $\Delta S=2$ four-fermion operator is
augmented by four additional operators.
Our aim in the present work is to provide fully controlled results for
the corresponding additional mixing matrix elements.

Calculations of such matrix elements using lattice QCD have a fairly long history.
Initial results were obtained starting in the late 1990s in the quenched 
approximation~\cite{Allton:1998sm,Donini:1999nn,Babich:2006bh}.
Then, in 2012, first results with unquenched light quarks were 
presented by the ETM~\cite{Bertone:2012cu} and RBC-UKQCD 
collaborations~\cite{Boyle:2012qb}.
These calculations used, respectively, twisted-mass and domain-wall lattice
fermions. Both performed the matching of lattice and continuum operators
using non-perturbative renormalization (NPR)~\cite{Martinelli:1994ty}
and the RI-MOM scheme.
The results for all four BSM $B$-parameters were consistent between the
two calculations.

In 2013, we presented results from a first calculation of the BSM $B$-parameters using
improved staggered fermions and one-loop perturbative
matching of lattice and continuum operators~\cite{Bae:2013tca}.
Our results disagreed significantly for two of the four $B$-parameters
with those from Refs.~\cite{Bertone:2012cu,Boyle:2012qb}.
In 2014, we discovered a minor error in our analysis that changed our
results by $\sim 5\%$.
We also extended the range of lattice ensembles studied, so that the
continuum and chiral extrapolations were better controlled.
Preliminary results correcting the analysis 
and incorporating the new ensembles were presented in Ref.~\cite{Jang:2014aea}.
The discrepancy with Refs.~\cite{Bertone:2012cu,Boyle:2012qb} remained
at about the $3\sigma$ level for two of the $B$-parameters.

The purpose of the present paper is provide a detailed description
of our calculation along with our final results. In fact, these results are
very close to the preliminary numbers presented in Ref.~\cite{Jang:2014aea},
but there are many details not provided in either Ref.~\cite{Bae:2013tca}
or~\cite{Jang:2014aea} that we present here.
A further motivation for this work is provided by the recent results
from the ETM and RBC-UKQCD collaborations, presented in
Refs.~\cite{Carrasco:2015pra} and at Lattice 2015 \cite{Hudspith:2015wev,RBC-UKQCDinprep},
respectively.
The former work
(which extends the $N_f=2$ simulations of Ref.~\cite{Bertone:2012cu}
to $N_f=2+1+1$)
essentially confirms the earlier results of Ref.~\cite{Bertone:2012cu},
and thus continues to disagree with our results.
The latter calculation, Ref.~\cite{Hudspith:2015wev}, presents an
investigation of the origin of the discrepancies by repeating their
computation with a second lattice spacing and performing the
renormalization with various schemes, including RI-SMOM schemes with
non-exceptional kinematics~\cite{Sturm:2009kb}.  Although the
discretization artifacts are found to be larger than previously
anticipated, the most important effects come from the renormalization
procedure. The preliminary results of RBC-UKQCD with the new SMOM
schemes are in approximate agreement with those presented here
\cite{RBC-UKQCDinprep}.
Given this complicated and confusing situation, it is important to have a clear
description of the details of all the calculations.

Our work relies on several auxiliary theoretical calculations.  For
the chiral extrapolations we need results from SU(2) staggered chiral
perturbation theory (SChPT), and these are provided in
Ref.~\cite{Bailey:2012wb}.  We also need to know how to set up the
calculation using staggered fermions (i.e. dealing with the extra
valence tastes) as well as one-loop matching factors. These results
are provided in Refs.~\cite{Kim:2011pz,Kim:2014tda}.  Finally, we need
to evolve matrix elements using the continuum renormalization group
for $\Delta S=2$ operators.  The required two-loop anomalous
dimensions were calculated in Ref.~\cite{Buras:2000if}, and some
additional technical details are worked out in
Ref.~\cite{Kim:2014tda}.

This paper is organized as follows.
In Sec.~\ref{sec:formalism}, we describe the basis of $\Delta S=2$
four-quark operators that we use, and the corresponding B-parameters
and gold-plated combinations.
In Sec.~\ref{sec:data}, we describe the details of the
lattice calculation.
We next turn to the analysis.
Sec.~\ref{sec:fit-su2} explains how we extrapolate valence
quark masses to their physical values,
while Sec.~\ref{sec:conti-chi} describes the combined extrapolation
to the continuum limit and to physical sea-quark masses.
We present our final results and error budget in 
Sec.~\ref{sec:final}, and compare these to the above-mentioned
results that use other fermion discretizations in Sec.~\ref{sec:conclude}.
%

\section{$\Delta S=2$ Four-quark Operators and Bag Parameters}
\label{sec:formalism}

We use the operator basis (Buras's basis) of Ref.~\cite{Buras:2000if},
in which the $\Delta S=2$ four-quark operators are
%
%
\begin{align}
\label{eq:ops}
\begin{split}
 {Q}_{1} &=
 [\bar{s}^a \gamma_\mu (1-\gamma_5) d^a] 
 [\bar{s}^b \gamma_\mu (1-\gamma_5) d^b]  \,, \\
 {Q}_{2} &=
 [\bar{s}^a (1-\gamma_5) d^a] 
 [\bar{s}^b (1-\gamma_5) d^b]   \,, \\
 {Q}_{3} &=
 [\bar{s}^a \sigma_{\mu\nu}(1-\gamma_5) d^a] 
 [\bar{s}^b \sigma_{\mu\nu} (1-\gamma_5) d^b] \,,  \\
 {Q}_{4} &=
 [\bar{s}^a (1-\gamma_5) d^a] 
 [\bar{s}^b (1+\gamma_5) d^b]\,,  \\
 {Q}_{5} &=
 [\bar{s}^a \gamma_\mu (1-\gamma_5) d^a] 
 [\bar{s}^b \gamma_\mu (1+\gamma_5) d^b] \,.
\end{split}
\end{align}
Here the operators have been written in Euclidean space,
with $a$ and $b$ being color indices.
$Q_1$ is the operator corresponding to $B_K$, while
$Q_{2,3,4,5}$ are the BSM operators.\footnote{%
This basis is complete in four dimensions aside from the
need to add the parity conjugates of $Q_1-Q_3$. 
We do not consider these additional operators, however,
since they have the same positive parity parts as $Q_1-Q_3$,
and the matrix element we consider picks out the positive
parity parts.}

The hadronic matrix elements of the $\Delta S=2$ four-quark 
operators can be parametrized by so-called kaon bag parameters 
(or $B$-parameters).
These are conventionally defined by
\begin{align}
B_K &= 
 \frac{\langle \overline{K}_0 \vert Q_1 \vert K_0 \rangle}
      {\frac{8}{3} \langle \overline{K}_0 \vert 
        \overline{s}\gamma_\mu \gamma_5 d\vert 0 \rangle
      \langle 0 \vert \bar{s} 
      \gamma_\mu \gamma_5 d \vert K_0 \rangle}
\\
\label{eq:def-B_i}
 B_j &= 
 \frac{\langle \overline{K}_0 \vert Q_j \vert K_0 \rangle}
      {N_j \langle \overline{K}_0 \vert \overline{s}\gamma_5 d\vert 0 \rangle
      \langle 0 \vert \bar{s} \gamma_5 d \vert K_0 \rangle}\,,
\end{align}
where $j=2-5$, and
\begin{align}
  (N_2,\ N_3,\ N_4,\ N_5) = ( 5/3, \ 4, \ -2, \ 4/3 )
\end{align}
are factors arising in the vacuum saturation approximation.
In the following, we will often refer collectively to 
``the $B_i$'', and this will indicate all five of the $B$-parameters,
i.e. the index $i$ runs over $i=K,2,3,4,5$.

In our lattice calculation, we find it more convenient to evaluate
the B-parameters rather than the corresponding matrix elements,
$\langle \overline{K}_0 \vert Q_i \vert K_0 \rangle$.
This avoids the need to determine the overlap of our sources with
the $\overline{K}_0$ and $K_0$ states, reduces the dependence
on the scale, $a$, since the $B$-parameters are dimensionless,
cancels some of statistical and systematic errors, 
and simplifies chiral expansions,
since the staggered chiral perturbation theory (SChPT) 
expressions are simpler~\cite{Bailey:2012wb}.

We also make extensive use of
``gold-plated'' combinations of the B-parameters.
These are combinations chosen to be free of chiral logarithms
at  next-to-leading order (NLO) in
SU(2) chiral perturbation theory~\cite{Bailey:2012wb}:
\begin{align}
\label{eq:gold-comb}
\begin{split}
  G_{21} &\equiv \frac{B_2}{B_K}, \quad
  G_{23} \equiv \frac{B_2}{B_3}, \\
  G_{24} &\equiv B_2 \cdot B_4, \quad 
  G_{45} \equiv \frac{B_4}{B_5}.
\end{split}
\end{align}
In this paper, the subindex $i$ of the $G_i$ runs over 
$i=21,23,24,45$.

As described below, it turns out that the combined extrapolation
in $a^2$ and sea-quark masses is much better controlled
for the $G_i$ and $B_K$ than for $B_{2-5}$.
Thus our final results for the BSM $B$-parameters are obtained
using $B_K$ and the $G_i$ in the following way:
\begin{align}
\label{eq:b-from-gold}
\begin{split}
  B^G_2 &= B_K \cdot G_{21} \,, \\
  B^G_3 &= B_K \cdot \frac{G_{21}}{G_{23}} \,, \\
  B^G_4 &= \frac{ G_{24} }{ B_K \cdot G_{21} } \,, \\
  B^G_5 &= \frac{ G_{24} }{ B_K \cdot G_{21} \cdot G_{45}} \,.
\end{split}
\end{align}
The superscript $G$ indicates that we use gold-plated
combinations to reconstruct the $B$-parameters.

\section{Lattices and Measurements}
\label{sec:data}

%
\begin{table}[htbp]
\caption{MILC ensembles used in our numerical study.
  Here ``ens'' represents the number of gauge configurations,
``meas'' is the number of measurements per configuration,
  and ID is a label. $am_\ell$ and $am_s$ are, respectively,
the light and strange sea quark masses in lattice units. The values
of $a$ are nominal.
  \label{tab:milc-lat}}
\begin{ruledtabular}
\begin{tabular}{c  c  c  c  l }
$a$ (fm) & $am_l/am_s$ & \ \ size & ens $\times$ meas  & ID \\
\hline
0.12  & 0.03/0.05    & $20^3 \times 64$  & $564 \times 9$ & C1 \\
0.12  & 0.02/0.05    & $20^3 \times 64$  & $486 \times 9$ & C2 \\
0.12  & 0.01/0.05    & $20^3 \times 64$  & $671 \times 9$ & C3 \\
0.12  & 0.01/0.05    & $28^3 \times 64$  & $274 \times 8$ & C3-2 \\
0.12  & 0.007/0.05   & $20^3 \times 64$  & $651 \times 10$& C4 \\
0.12  & 0.005/0.05   & $24^3 \times 64$  & $509 \times 9$ & C5 \\
\hline
0.09  & 0.0062/0.031 & $28^3 \times 96$  & $995 \times 9$ & F1 \\
0.09  & 0.0031/0.031 & $40^3 \times 96$  & $959 \times 9$ & F2 \\
0.09  & 0.0093/0.031 & $28^3 \times 96$  & $949 \times 9$ & F3 \\
0.09  & 0.0124/0.031 & $28^3 \times 96$  & $1995\times 9$ & F4 \\
0.09  & 0.00465/0.031& $32^3 \times 96$  & $651 \times 9$ & F5 \\
0.09  & 0.0062/0.0186& $28^3 \times 96$  & $950 \times 9$ & F6 \\
0.09  & 0.0031/0.0186& $40^3 \times 96$  & $701 \times 9$ & F7 \\
0.09  & 0.00155/0.031& $64^3 \times 96$  & $790 \times 9$ & F9 \\
\hline
0.06  & 0.0036/0.018 & $48^3 \times 144$ & $749 \times 9$ & S1 \\
0.06  & 0.0025/0.018 & $56^3 \times 144$ & $799 \times 9$ & S2 \\
0.06  & 0.0072/0.018 & $48^3 \times 144$ & $593 \times 9$ & S3 \\
0.06  & 0.0054/0.018 & $48^3 \times 144$ & $582 \times 9$ & S4 \\
0.06  & 0.0018/0.018 & $64^3 \times 144$ & $572 \times 9$ & S5 \\
\hline
0.045 & 0.0028/0.014 & $64^3 \times 192$ & $747 \times 1$ & U1 \\
\end{tabular}
\end{ruledtabular}
\end{table}

We use the MILC ensembles listed in Table~\ref{tab:milc-lat}.
These are generated with $N_f = 2+1$ flavors of staggered fermions
using the ``asqtad'' fermion action.
Details of the configuration generation are given in
Ref.~\cite{Bazavov:2009bb}.
To convert our data to physical units, 
we use the values of $r_1/a$ obtained by the MILC 
collaboration.~\cite{Bazavov:2009bb, milc-private}, 
and set $r_1 =0.3117(22)\textrm{ fm}$,
following Refs.~\cite{milc-private, Bailey:2012rr}.\footnote{%
Some values of $r_1/a$ are updated compared to Ref.~\cite{Bazavov:2009bb};
these are
F2: $3.6987$, F3: $3.7036$, F4: $3.7086$, F5:$3.6993$, F7: $3.7000$,
F9: $3.6984$, S4: $5.2825$, S5: $5.2836$~\cite{milc-private}.}
We stress that the values of $a$ listed in the table are nominal.
The actual values (determined from $r_1/a$) differ slightly from the
nominal values, and it is the former that we use in our analysis.
In the following, we sometimes use MILC terminology and
refer to the sets of ensembles with nominal
lattice spacings of $a=0.12\;$fm, $0.09\;$fm, $0.06\;$fm and $0.045\;$fm
as coarse, fine, superfine and ultrafine lattices, respectively.

Compared to the results presented in Ref.~\cite{Bae:2013tca},
the additional ensembles are F6, F7, F9 and S5.
These additions significantly improve the reliability of
the chiral extrapolations, as we now explain.
On all ensembles except F6 and F7, the strange sea quark masses
lie close to, but not exactly at, the physical value.
Adding in F6 and F7, which have lighter strange sea quarks,
allows us to correct for the offset.
Adding S5 ensures that on both fine and
superfine lattices the average up/down
sea quark mass, $m_\ell$, ranges down to $\approx m_s^{\rm phys}/10$,
so that the chiral extrapolation is relatively short.
Finally, adding in F9 provides us with a light sea quark mass,
$m_\ell \approx m_s^{\rm phys}/20$, that lies much closer to the
physical value.

We use HYP-smeared staggered fermions~\cite{Hasenfratz:2001hp} as
valence quarks.
Parameters for the HYP smearing are chosen to remove
$\mathcal{O}(a^2)$ taste-symmetry breaking at tree level
\cite{Lee:2002ui}.
We use 10 different valence quark masses on each lattice:
\begin{align}
\label{eq:val_q_masses}
  m_{x},\ m_{y} = m^\text{nom}_s \times \frac{n}{10} 
  \qquad \text{ with } n=1,2,3,\ldots,10
\,,
\end{align}
where $m^\text{nom}_s$ is the nominal strange quark mass given
in Table \ref{tab:val-qmass}. We have labeled the valence masses
$m_x$ and $m_y$, the former corresponding to the valence $d$ quark
and the latter to the valence $s$ quark.

As explained in the next section,
$m_x$ and $m_y$ will be extrapolated to their physical values,
$m_d^{\rm phys}$ and $m_s^{\rm phys}$, respectively.
To determine these physical values on each ensemble use
the same method as in Ref.~\cite{Bae:2010ki}. 
First, the flavor non-singlet $y\bar{y}$ ``pion'' mass is extrapolated
until is equals $M_{ss,\rm phys}=0.6858(40) \text{ GeV}$, which is the ``physical'' value
determined in Ref.~\cite{Davies:2009tsa}.
This determines $m_s^{\rm phys}$.
Second, $m_x$ is extrapolated (with $m_y$ at its now-determined physical value)
such that the $x\bar y$ ``kaon'' has a mass equal to that of the physical $K_0$.
These extrapolations are done separately on each ensemble.
For illustration, we show in 
Table~\ref{tab:val-qmass} the resulting physical values
(as well as the valence masses we use in the simulations)
for the ensembles having $m_\ell/m_s=1/5$. 
We see that our lightest valence quark masses are roughly twice
$m_d^{\rm phys}$, while our heaviest lie somewhat below 
$m_s^{\rm phys}$.
%
%
%
\begin{table}[htbp]
\caption{Physical values of valence quark masses on representative ensembles,
in lattice units.
For comparison, we also show the range of  valence masses used in simulations.
\label{tab:val-qmass}}
\begin{ruledtabular}
\begin{tabular}{ccccc}
Ensemble & $a m_d^\text{phys}$ & $a m_s^\text{phys}$  & $a m_x$ and $a m_y$ \\ \hline
C3       & 0.00213(2) & 0.05204(5) & $0.005$--$0.05$   \\ 
F1       & 0.00146(2) & 0.03542(5) & $0.003$--$0.03$   \\ 
S1       & 0.00104(1) & 0.02372(3) & $0.0018$--$0.018$   \\ 
U1       & 0.00076(1) & 0.01693(3) & $0.0014$--$0.014$  \\
\end{tabular}
\end{ruledtabular}
\end{table}

We calculate the valence $x \bar x $ ``pion" and $x \bar y$ ``kaon" masses in
standard fashion using the same wall sources as described below. 
The statistical errors on these results are very small.
In Table~\ref{tab:pion-mass} we quote some representative values to indicate
the range of pion masses in physical units. Note that $m_\pi({\rm val,max})$ is
the mass of the heaviest pion that we use in our chiral extrapolation to the
physical valence $d$ quark. This extrapolation is discussed in the following section.
We also include values for the lightest sea-quark pion for the fine, superfine and
ultrafine lattices, as well as for the coarse ensemble that we use to study finite-volume effects.

%
%
\begin{table}[htbp]
\caption{ Valence and sea pion masses (in GeV) on
  representative ensembles.  $m_\pi({\rm val,  min})$ and
  $m_\pi({\rm val, max})$ are the minimum and maximum valence pion masses
  used in our valence chiral extrapolation.  
  The values for these quantities on other coarse, fine and superfine ensembles
  are very similar to those on ensembles C3, F9 and S5, respectively.
  For the fine, superfine and ultrafine lattices, we show the sea quark
pion mass on the ensemble with the smallest value of this quantity.
 For the coarse lattices, we pick the ensemble used to estimate finite-volume
effects.
\label{tab:pion-mass}}
\begin{ruledtabular}
\begin{tabular}{cccc}
Ensemble & $m_\pi({\rm val, min})$ & $m_\pi({\rm val, max})$ & $m_\pi({\rm sea})$ \\ \hline
C3       & 0.222 & 0.430 & 0.372 \\
F9       & 0.206 & 0.401 & 0.174 \\
S5       & 0.195 & 0.379 & 0.222 \\
U1       & 0.206 & 0.397 & 0.316 \\
\end{tabular}
\end{ruledtabular}
\end{table}

We use essentially the same  methodology for calculating the BSM
$B$-parameters as we employed in the calculation of $B_K$ 
in Ref.~\cite{Bae:2010ki}. Thus we give only a brief discussion here,
while for $B_K$ we refer to Ref.~\cite{Bae:2010ki}.
In terms of lattice operators, the BSM B-parameters are
\begin{align}
  &B_j(t) = 
  \frac{2 \langle \overline{K}_{P1}^0 \vert 
        z_{jk} Q^\text{Lat}_k(t) \vert K_{P2}^0 \rangle}
       {N_j \langle \overline{K}_{P1}^0 \vert 
         z_P \mathcal{O}_P^\text{Lat}(t) \vert 0 \rangle \langle 0 \vert 
         z_P \mathcal{O}_P^\text{Lat}(t) \vert K_{P2}^0 \rangle}
\,,
\end{align}
where $Q^\text{Lat}_k$ are lattice four-fermion operators and 
$\mathcal{O}_P^\text{Lat}$ is the taste $\xi_5$ pseudoscalar bilinear.
$z_{jk}$ and $z_P$ are one-loop matching factors that 
convert lattice operators to their continuum counterparts,
the latter defined in the $\overline{\text{MS}}$ scheme using naive dimensional 
regularization (NDR).
%
We use the mean-field improved 
lattice operators defined in
Refs.~\cite{Kim:2011pz,Kim:2014tda}.
The one-loop matching is quite involved as one must ensure that
the continuum basis is extended to $d=4-2\epsilon$ dimensions using
the same definition of evanescent operators as in Ref.~\cite{Buras:2000if}.
The matching factors have been worked out and described in detail in
Ref.~\cite{Kim:2014tda}, building on the earlier work
of Ref.~\cite{Kim:2011pz}, and we do not repeat them here.
They depend on the renormalization scale $\mu$ of the continuum operator
and on $\alpha_s$.
The latter is chosen to be in the $\overline{\text{MS}}$ scheme
and is evaluated at the same scale $\mu$.
In our initial matching we take $\mu=1/a$ and then evolve the results
in the continuum to a common renormalization scale.
In the numerical evaluation of the matching coefficients we use
four loop running to determine $\alpha_s(\mu)$, using as input
$\alpha_s(M_Z)=0.118$.

To produce the kaons and antikaons, we place U(1)-noise 
wall sources on time slices $t_1$ and $t_2$, with $t_2>t_1$.
These produce taste $\xi_5$ kaons and anti-kaons having zero 
spatial momenta.
The four-quark operators are placed between the sources
at time $t$ (i.e. $t_1 < t < t_2$).
When $t$ is far enough from the sources, so that excited state
contamination is small, the three-point correlators should be independent
of $t$, and can be fit to a constant.
To determine the fit range, we use the two-point correlator from 
the wall-source to the taste $\xi_5$ axial current.
From the effective mass plot for this correlator, 
we find the distance from the source, $t_L$, for which
the contamination from 
excited states becomes negligibly small.
Then we fit from $t=t_1 + t_L$ to 
$t = t_1 + t_R = t_2 - t_L -1$
(which is a symmetrical range since our operators extend over the
two time slices $t$ and $t+1$).
Our choices of $t_L$ and $t_R$  are given in Table~\ref{tab:g-fit-range}.
Note that we choose $\Delta t=t_2-t_1$ to be less than half of the time
extent of the lattice to avoid ``around the world" contributions.
Further details concerning sources and time ranges is given in 
Ref.~\cite{Bae:2010ki}.

%
%
%
\begin{table}[htbp]
\caption{Choices for the wall source separation, $\Delta t = t_2 - t_1$,
and its ratio to the temporal length of the lattices, $T$,
as well as the parameters determining the fitting range.
  \label{tab:g-fit-range}}
\begin{ruledtabular}
\begin{tabular}{ l  l  l l l l}
lattice spacing & $\Delta t$ & $\Delta t / T$ & $t_L$ & $t_R$ & $t_L$ (fm) \\ 
\hline
$0.12\;$fm &  26  & 0.41  & 10 & 15 & 1.19 \\
$0.09\;$fm &  40  & 0.42  & 14 & 25 & 1.18 \\
$0.06\;$fm &  60  & 0.42  & 22 & 37 & 1.29 \\
$0.045\;$fm &  80  & 0.42  & 26 & 53 & 1.14 \\
\end{tabular}
\end{ruledtabular}
\end{table}

\begin{figure*}[tbhp]
  \subfigure[F1]{
    \label{fig:G_23:F1}
    \includegraphics[width=0.31\textwidth]{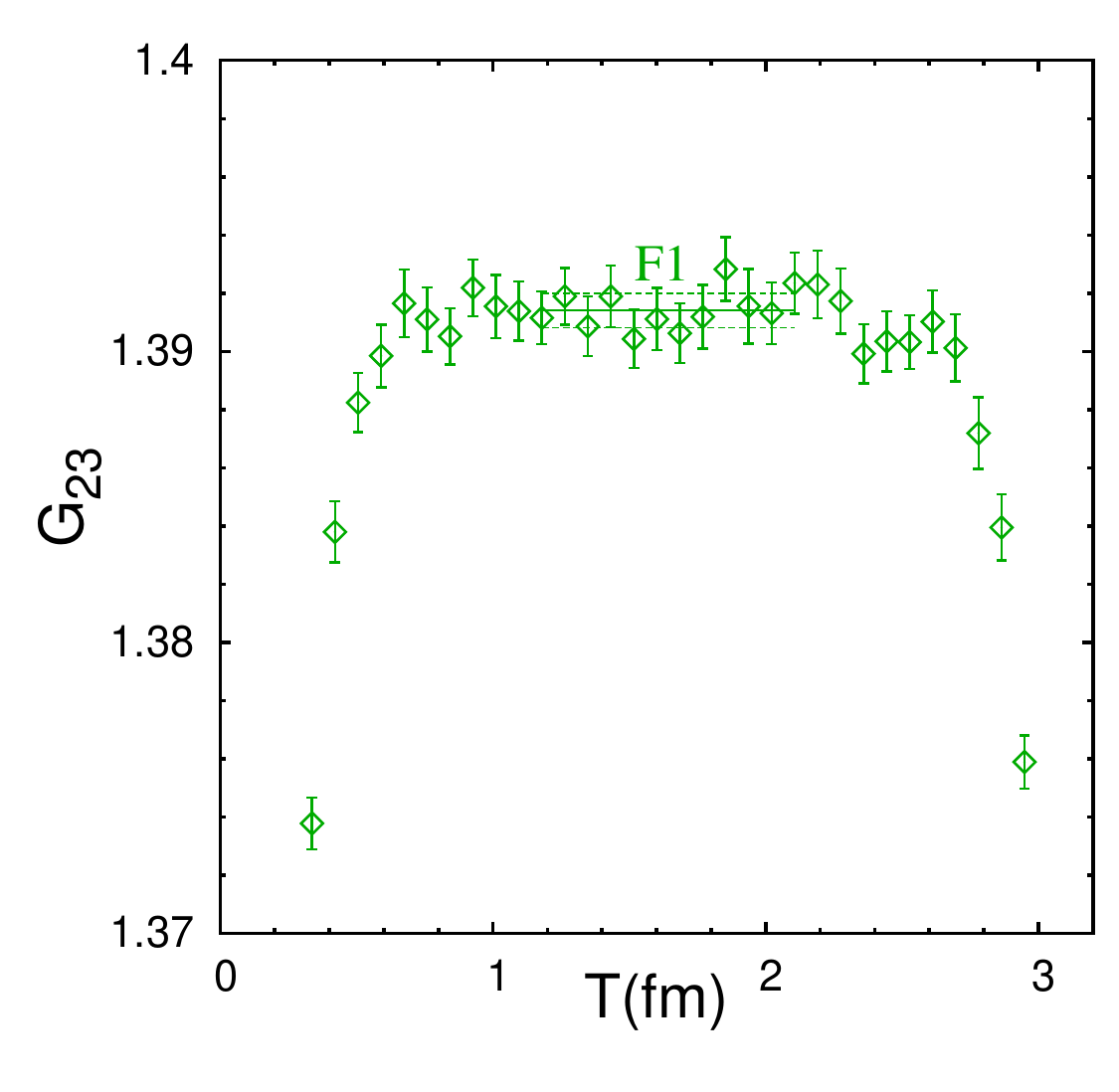}
  }
  \hfill
  \subfigure[S1]{
    \label{fig:G_23:S1}
    \includegraphics[width=0.31\textwidth]{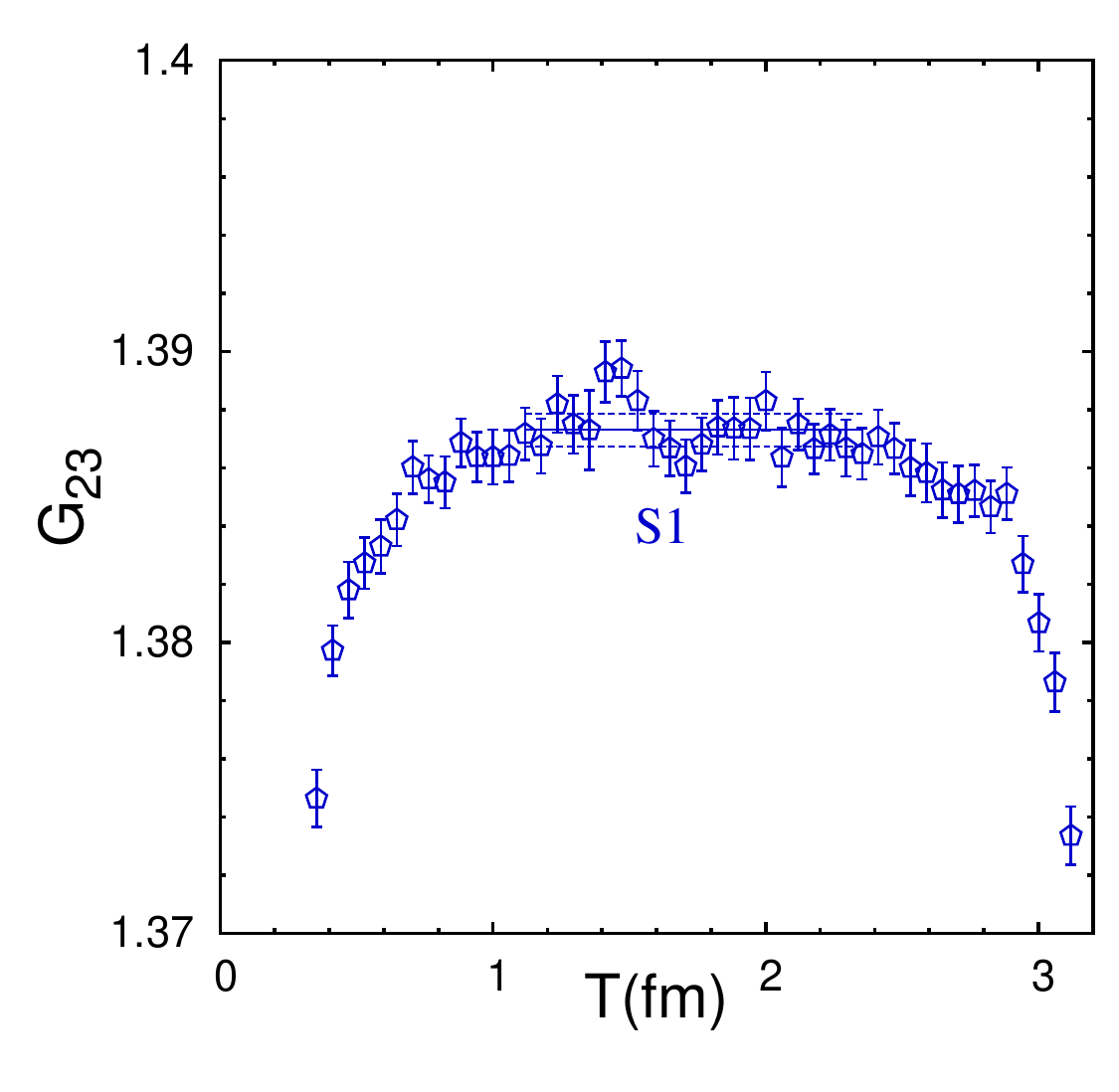}
  }
  \hfill
  \subfigure[U1]{
    \label{fig:G_23:U1}
    \includegraphics[width=0.31\textwidth]{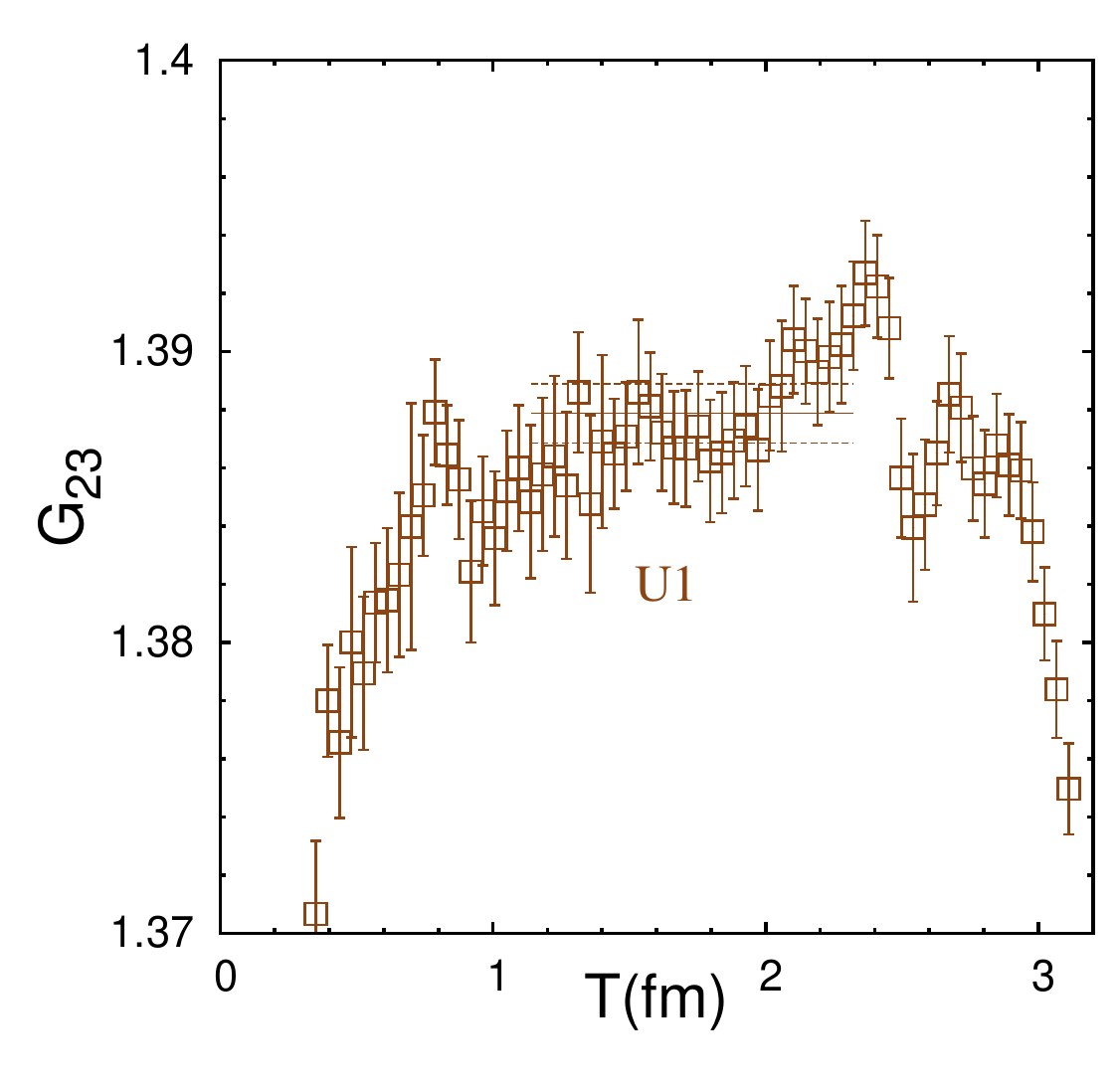}
  }    
  \caption{ $G_{23}$ (evaluated at renormalization scale $\mu=1/a$
  as a function of $T = t-t_1$. Green
    diamonds show on the F1 ensemble with $(am_x, am_y)
    = (0.003, 0.03)$. Blue pentagons are results from the S1 ensemble
    with $(am_x, am_y) = (0.0018, 0.018)$.  Brown squares are results
    on the U1 ensemble with $(am_x, am_y) = (0.0014, 0.014)$. 
    The fit ranges (and resulting central values and error bands) are shown by
    the horizontal lines.}
  \label{fig:G_23}
\end{figure*}

\begin{figure}[tbhp]
  \begin{center}
    \includegraphics[width=0.40\textwidth]{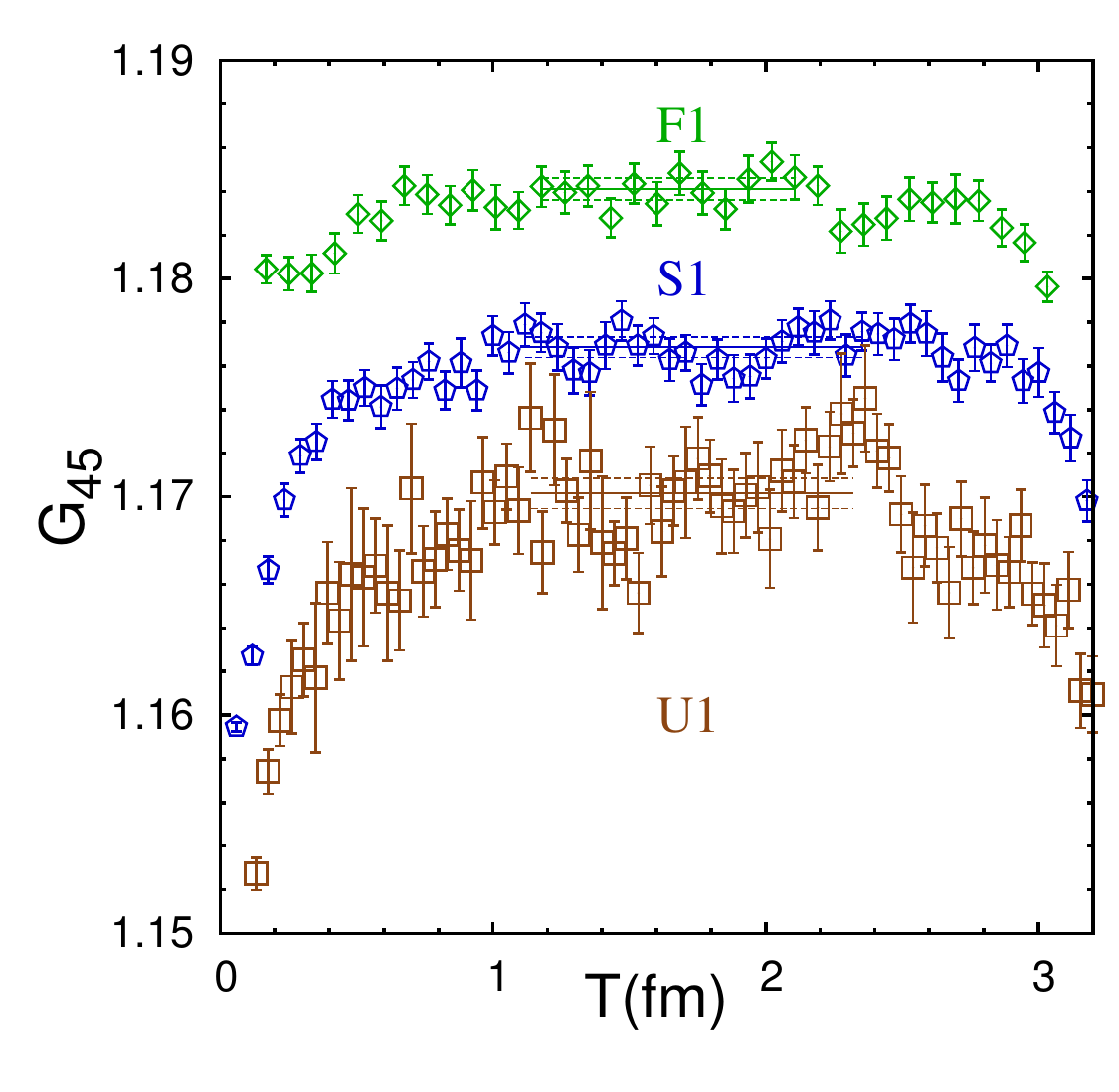}    
  \end{center}
  \caption{ $G_{45}$ as a function of $T = t-t_1$ at
    $\mu=1/a$. The convention for symbols is as in Fig.~\ref{fig:G_23}.}
  \label{fig:G_45}
\end{figure}

The plateaus resulting from the above-described procedure are
reasonable. Examples are shown for the
gold-plated combinations $G_{23}$ and $G_{45}$
in Figs.~\ref{fig:G_23} and \ref{fig:G_45}.
Here we show cases with light valence quark masses 
($m_x/m_y = 1/10$ with $m_l/m_s=1/5$)
for which the statistical errors are larger.
The fits to a constant  are performed ignoring correlations 
between time slices (diagonal approximation for the covariance
matrix) in order to avoid instabilities due to the small 
eigenvalues of the covariance matrix~\cite{Jang:2011fp}.
Fitting errors are estimated using the jackknife method.

To increase statistics, we do multiple measurements
on each configuration.
For each measurement, the source position $t_1$ is chosen 
randomly, with $t_2$ determined by $t_2 = t_1 + \Delta t$, 
where $\Delta t$ is the wall-source separation listed in 
Table~\ref{tab:g-fit-range}.
In addition, we use different random numbers for the wall sources for each measurement.
The number of measurements for each gauge configuration is 
listed in Table~\ref{tab:milc-lat}.

%
%
%
\begin{figure*}[htbp]
  \subfigure[$[P\times P\text{]}[P\times P\text{]}_I$]{
    \label{fig:autoc:1}
    \includegraphics[width=0.48\textwidth]{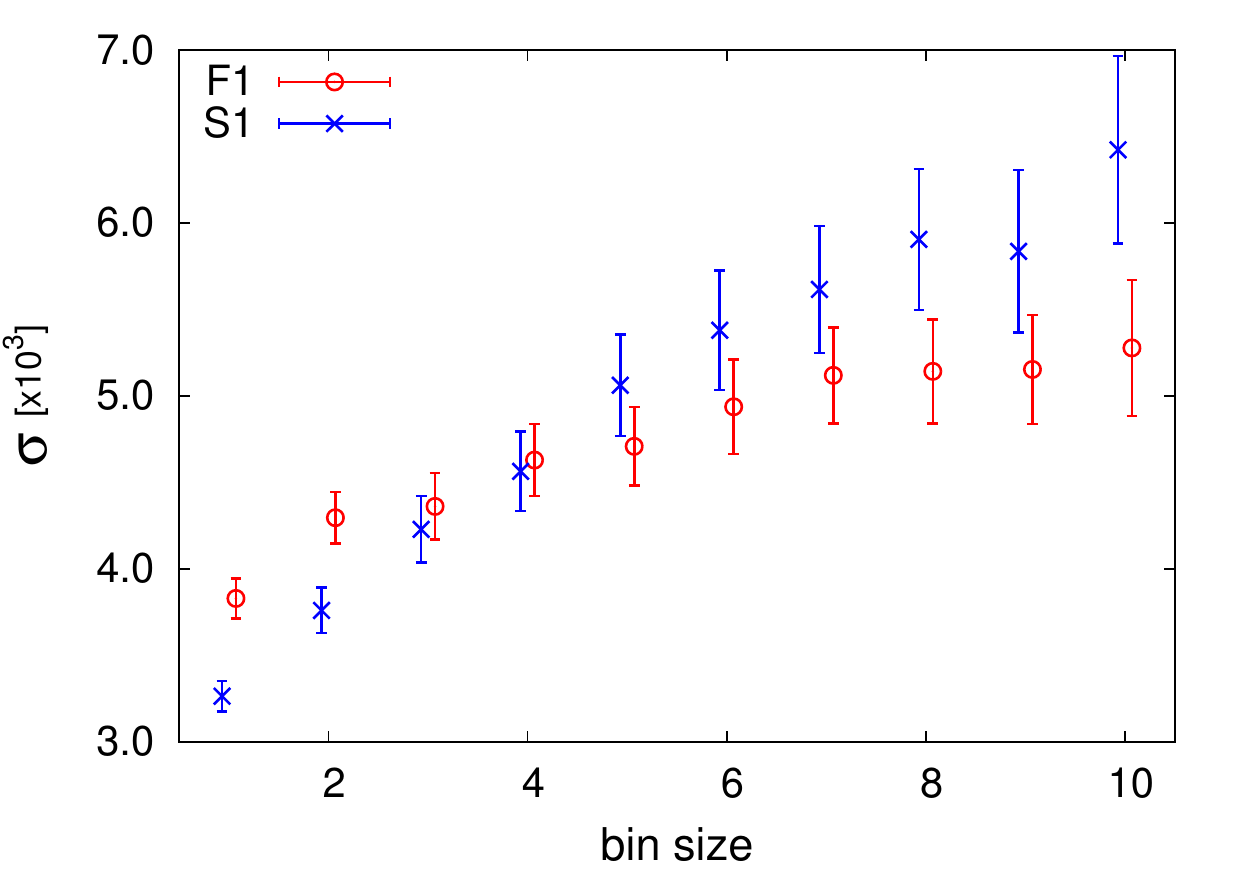}
  }
  \hfill
  \subfigure[$[P\times P\text{]}[P\times P\text{]}_{II}$]{
    \label{fig:autoc:2}
    \includegraphics[width=0.48\textwidth]{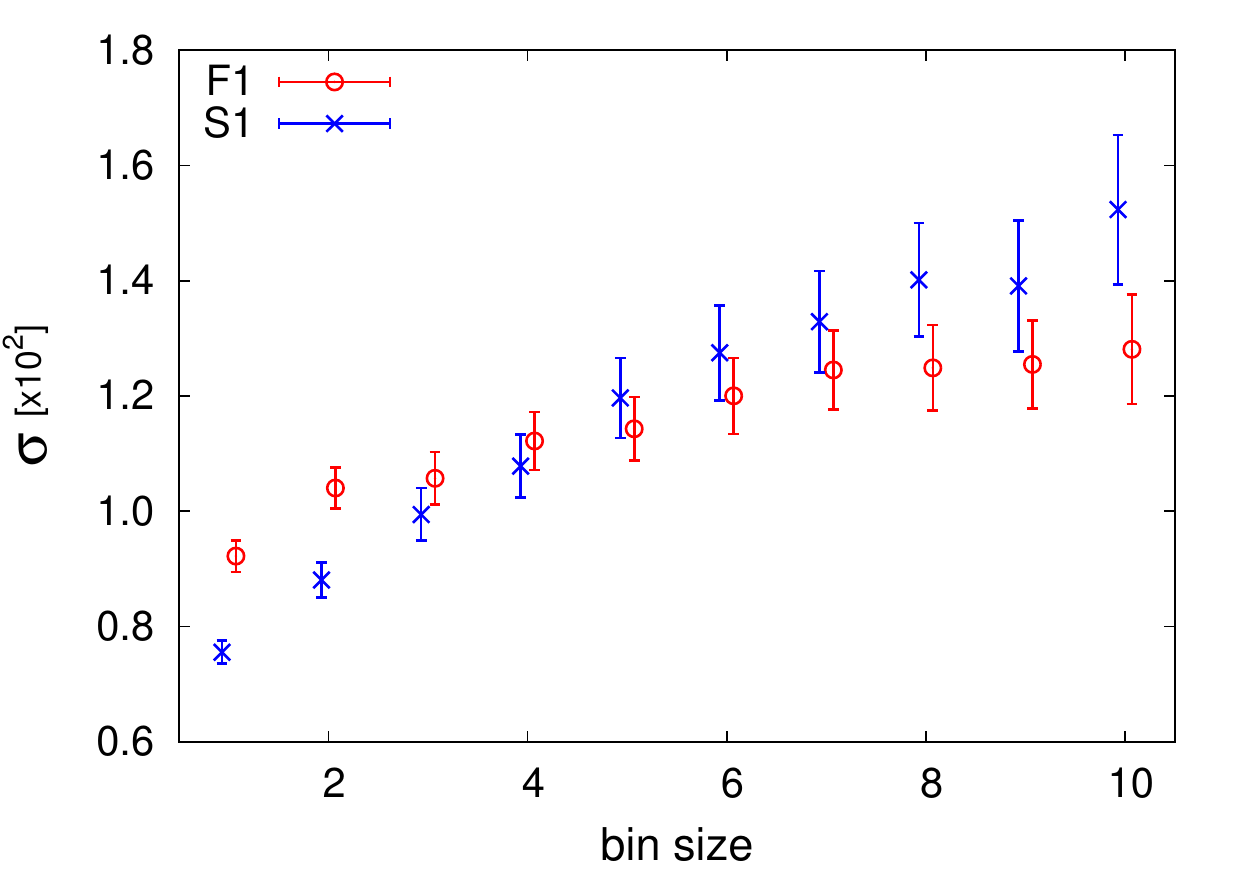}
  }
  \caption{ Statistical errors for bare three-point functions as a function of bin size.
    The operators are (a) $\mathcal{O}_{P1}^\textrm{Lat}$, and (b)
    $\mathcal{O}_{P2}^\textrm{Lat}$, respectively, at $T=20$ (F1) and
    $T=30$ (S1).  (Red) circles are the results on F1 ensemble, with
    $(am_x, am_y) = (0.003, 0.03)$; and (blue) crosses are the results
    on S1 ensemble, with $(am_x, am_y) = (0.0018, 0.018)$. }
  \label{fig:autocorr}
\end{figure*}

To study auto-correlations we bin adjacent lattices in the Markov chain
and study the dependence of the nominal statistical error on bin size.
Examples of the results are shown in Fig.~\ref{fig:autocorr}.
The notation for the operators used in this figure is explained in
Refs.~\cite{Kim:2011pz,Kim:2014tda}.
We find that the auto-correlations increase as the lattice 
spacing decreases.
As one can see from Fig.~\ref{fig:autocorr}, the auto-correlation 
effect is about 100\% for the MILC superfine lattice S1,
while it is about 25\% for the MILC fine lattice F1.
In order to greatly reduce the effects of auto-correlations, 
we use bins of size 5 throughout our analysis.

\section{Chiral Extrapolation}
\label{sec:fit-su2}
Our analysis follows the same steps as in Refs.~\cite{Bae:2013tca,Bae:2014sja}.
The first step is the chiral extrapolation of the valence 
quark masses to their physical values.
We extrapolate $m_x$ to 
the $m_d^\text{phys}$ for fixed $m_y$ using a fitting 
form based on SU(2) SChPT,
and then  extrapolate $m_y$ to $m_s^\text{phys}$.
For SU(2) ChPT to be valid, we require that $m_x \ll m_y$.
Hence, from the 10 valence quark masses listed in
Eq.~\eqref{eq:val_q_masses},
we take the lightest four for $m_x$ 
(e.g. $m_x = \{0.003,~0.006,~0.009,~0.012\}$ on the fine 
ensemble) and heaviest three for $m_y$ 
(e.g. $m_y = \{0.024,~0.027,~0.03\}$ on the fine ensemble).
In this way we satisfy $m_x \le m_y/2$.

We begin by considering the extrapolation in $m_x$, which we call the
``X fit".  The actual extrapolation is done in $X_P=m_{xx,P}^2$, which
is the squared mass of the $x\bar{x}$ valence pion with taste $\xi_5$
(i.e. the Goldstone pion).  For the physical value of this quantity we
take $X_P = 2 M_{K_0,\rm phys}^2-M_{ss, \rm phys}^2= (0.158 \; {\rm
  GeV})^2$.
At next-to-leading order (NLO) in SU(2) SChPT, the light valence quark mass 
dependence of the B-parameters has been worked out in
Ref.~\cite{Bailey:2012wb}, and is
\begin{align}
\label{eq:X-fit-NLO}
 B_i(\text{NLO}) 
  = c_1 F_0(i) + c_2 X, 
\end{align}
where $X \equiv \dfrac{X_P}{\Lambda_\chi^2}$ with
$\Lambda_\chi = 1\GeV$, the $c_j$ are coefficients to be determined, and
\begin{align}
\label{eq:f0}
 F_0(i) 
  &= 1 \pm \frac{1}{32\pi^2 f^2} \Bigg\{  \ell(X_I) 
   + (L_I-X_I)\tilde\ell(X_I)
 \nonumber \\
& \qquad \qquad \qquad \qquad \qquad
  - \frac{1}{16} \sum_B \ell(X_B) \Bigg\}
\,,
\end{align}
is the chiral logarithm.
Here $X_B$ ($L_B$) is the squared mass of the taste $B$, flavor non-singlet,
pion  composed of two light valence (sea) quarks:
$X_B = m^2_{xx,B}$ ($L_B = m^2_{ll,B}$). 
The functions $\ell(X)$ and $\tilde\ell(X)$ are chiral logarithms defined,
for example,  in Ref.~\cite{Bailey:2012wb}.
In Eq.~\eqref{eq:f0}, the plus sign applies for $i=K,2,3$, 
and the minus sign for $i=4,5$.

The NLO fitting function is not accurate enough to describe the precise and highly 
correlated data.
%
%
Hence, as in all our recent analyses~\cite{Bae:2011ff,Bae:2013tca,Bae:2014sja},
we add higher order terms to the fitting function:
\begin{align}
\label{eq:X-fit-NNNLO}
 B_j(\text{NNNLO}) 
  &= c_1 F_0(j) + c_2 X + c_3 X^2 + c_4 X^2 \ln^2(X)
    \nonumber \\
  &\quad +  c_5 X^2 \ln(X) + c_6 X^3.
\end{align}
The three terms $X^2, X^2 \ln^2(X)$ and 
$X^2 \ln(X)$ are the generic NNLO terms in continuum 
chiral perturbation theory.
We also add a single analytic NNNLO term proportional to $X^3$.
We use a similar fitting function for the X fits of gold-plated combinations, 
except that, by construction, there are no NLO chiral logarithms:
\begin{align}
\label{eq:X-fit-NNNLO-g}
 G_i(\text{NNNLO}) 
  &= c_1  + c_2 X + c_3 X^2 + c_4 X^2 \ln^2(X)
    \nonumber \\
  &\quad +  c_5 X^2 \ln(X) + c_6 X^3.
\end{align}
\sharpe{We have found that adding yet higher order terms in the
chiral expansion does not improve the fits to either the
$B_i$ or $G_i$.}

Since we have only four data points for the X fit, we use the
Bayesian method~\cite{Lepage:2001ym}, and place constraints
on the three higher-order fitting parameters $c_{4-6}$.
\sharpe{
Our prior information is that these coefficients are of $\mathcal{O}(1)$.
We thus first impose the constraints $c_{4-6} = 0 \pm 1$.
If the resulting fits have $\chi^2/{\rm d.o.f.} \lesssim 1$,
then we accept them. If not, we try the less restrictive constraints
$c_{4-6} = 0 \pm 2$. Again, we accept fits with 
$\chi^2/{\rm d.o.f.} \lesssim 1$,
but otherwise fit again using $c_{4-6} = 0 \pm 4$. 
In all cases this leads to
fits having $\chi^2/{\rm d.o.f.} \lesssim 1$.
In this discussion, the $\chi^2$ that is minimized is the augmented version:}
\begin{eqnarray}
  \chi^2_{\text{aug}}   & = \chi^2 + \chi^2_{\text{prior}}, \\
  \chi^2_{\text{prior}} & = {\displaystyle \sum_{i=4}^6 
  \frac{ (c_i - a_i)^2}{ \sigma_i^2} }
\,,
\end{eqnarray}
where we set $a_i = 0$ and \wlee{$\sigma_i = 1,2,4$}.
These fits are done using the full correlation matrix,
and have acceptable values of $\chi^2$.

Having determined the parameters $c_{1-6}$, we extrapolate 
the results to the physical point $m_x = m_d^\text{phys}$, 
and simultaneously remove (by hand) the lattice artifacts that lead to
taste symmetry breaking in pion masses.
Specifically, within the chiral logarithm $F_0(i)$ we set
$X_B$ and $L_I$ to their physical values, as 
explained in Ref.~\cite{Bae:2010ki}.
In this way we are using knowledge from SChPT to remove
a significant source of discretization errors.
Note that this correction applies to the $B_i$ but
not to the $G_j$, since the gold-plated
combinations have no chiral logarithms at NLO.

\begin{figure*}[tbhp]
  \subfigure[X-fit for $B_K$]{
    \label{fig:BK:xfit}
    \includegraphics[width=0.48\textwidth]
                    {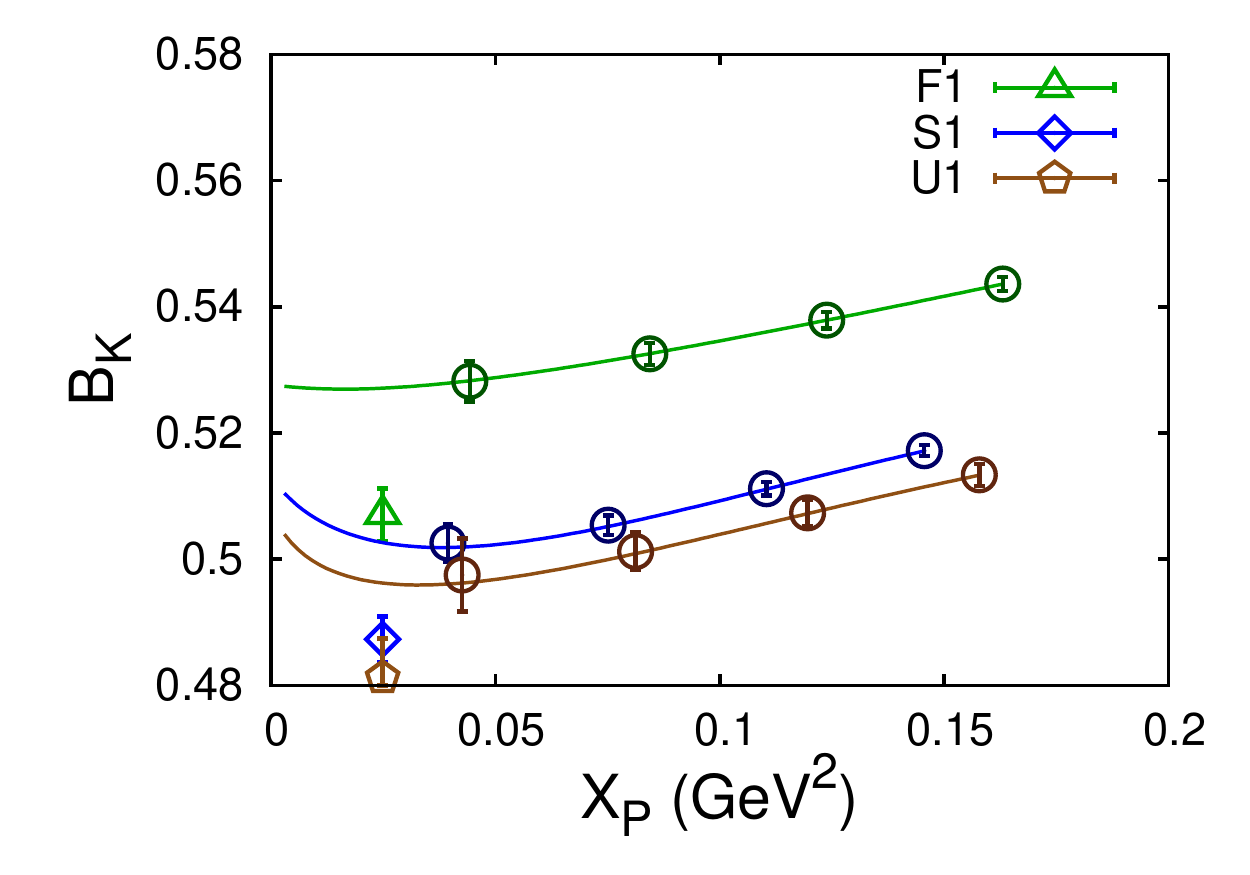}
  }
\hfill
  \subfigure[Y-fit for $B_{K}$]{
    \label{fig:BK:yfit}
    \includegraphics[width=0.48\textwidth]
                    {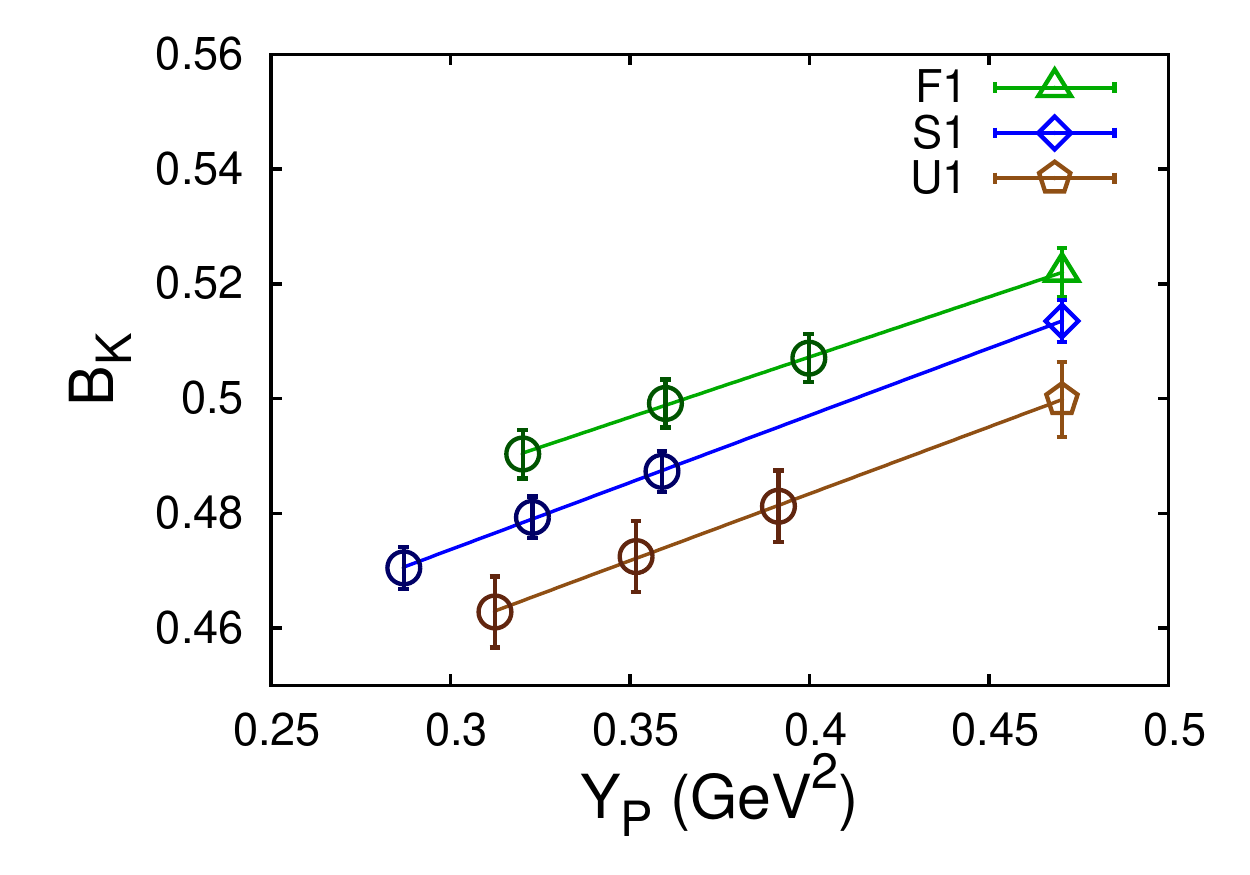}
  }
  \caption{ \subref{fig:BK:xfit} X fits and
    \subref{fig:BK:yfit} Y fits for $B_K$ evaluated at $\mu=1/a$ on the
    F1, S1 and U1 ensembles. The valence strange-quark masses are  $am_y = 0.03$,
    $0.018$ and $0.014$, respectively. 
    Lattice results are shown with circles (green, blue and brown for F1, S1, and U1, respectively)
    and are ordered vertically as shown in the legend.
    Extrapolated results are shown with [green] triangles (F1), 
    [blue] diamonds (S1) and [brown] pentagons (U1).
    For the X fit, the extrapolated results lie below the curves because of the removal
    of taste-breaking effects, as described in the text.}
  \label{fig:BK:XYfit}
\end{figure*}

\begin{figure*}[tbhp]
  \subfigure[X-fit for $G_{23}$]{
    \label{fig:G23:xfit}
    \includegraphics[width=0.48\textwidth]
                    {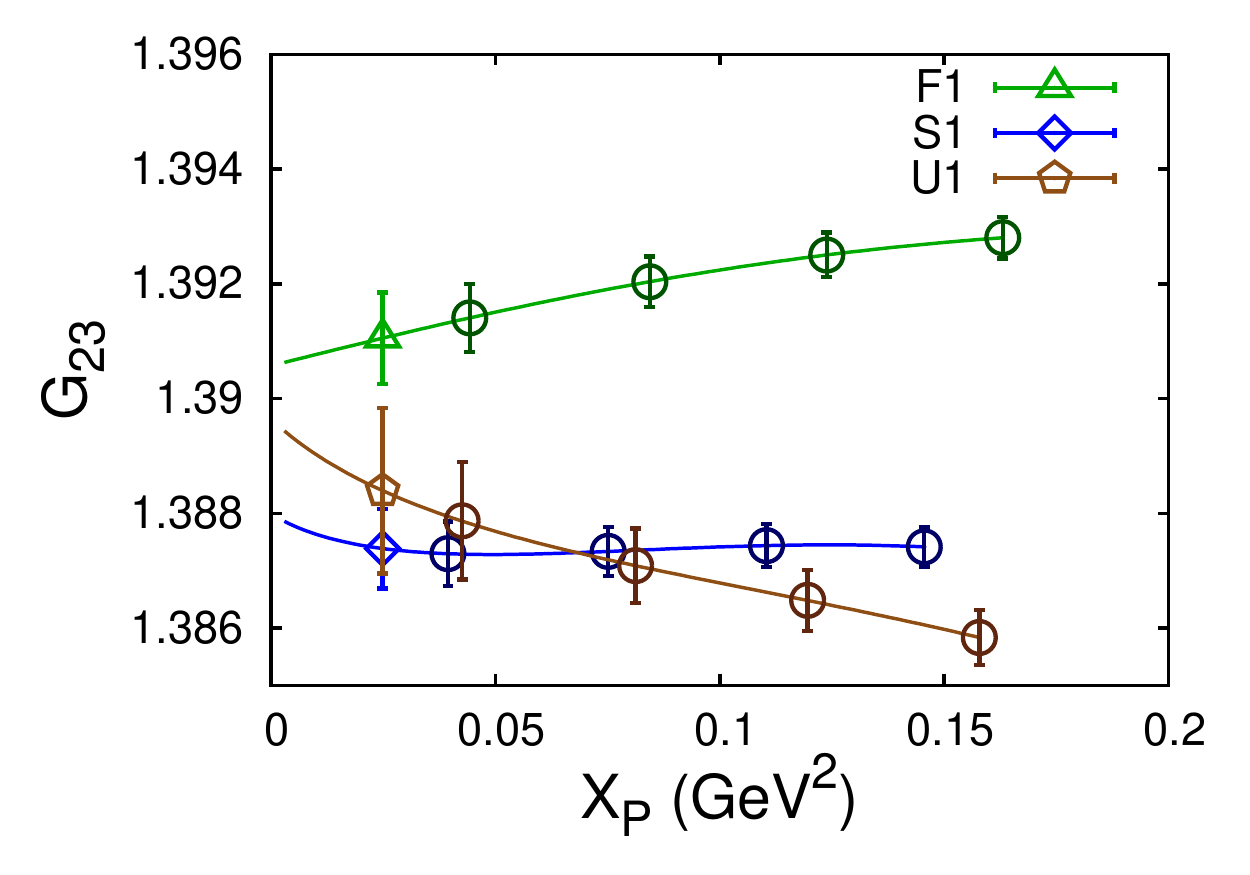}
  }
\hfill
  \subfigure[Y-fit for $G_{23}$]{
    \label{fig:G23:yfit}
    \includegraphics[width=0.48\textwidth]
                    {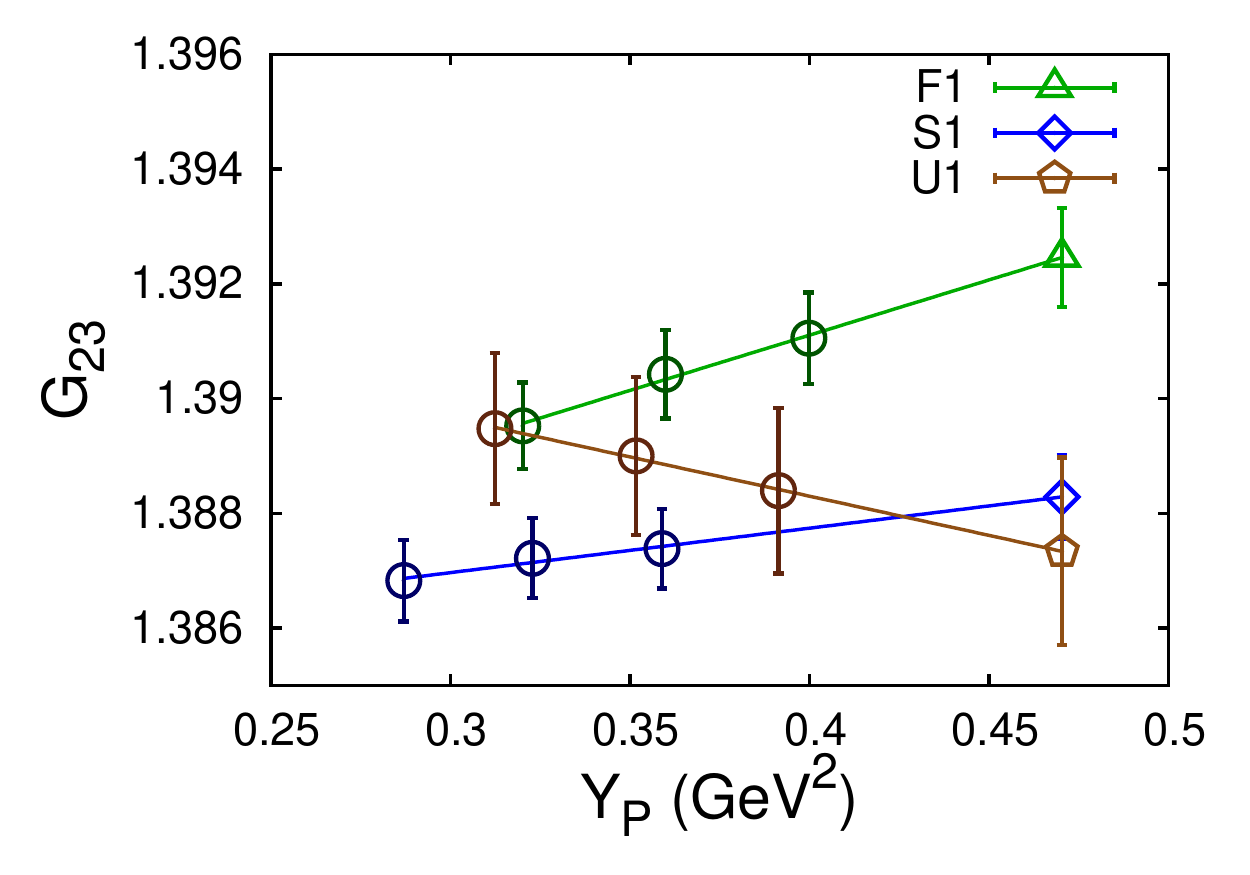}
  }
  \caption{\subref{fig:G23:xfit} X fits and
    \subref{fig:G23:yfit} Y fits for $G_{23}$. Notation as in
        Fig.~\ref{fig:BK:XYfit}, except that for the gold-plated combinations 
        there is no taste-breaking correction.}
    \label{fig:G23:XYfit}
\end{figure*}

\begin{figure*}[tbhp]
  \subfigure[X-fit for $G_{45}$]{
    \label{fig:G45:xfit}
    \includegraphics[width=0.48\textwidth]
                    {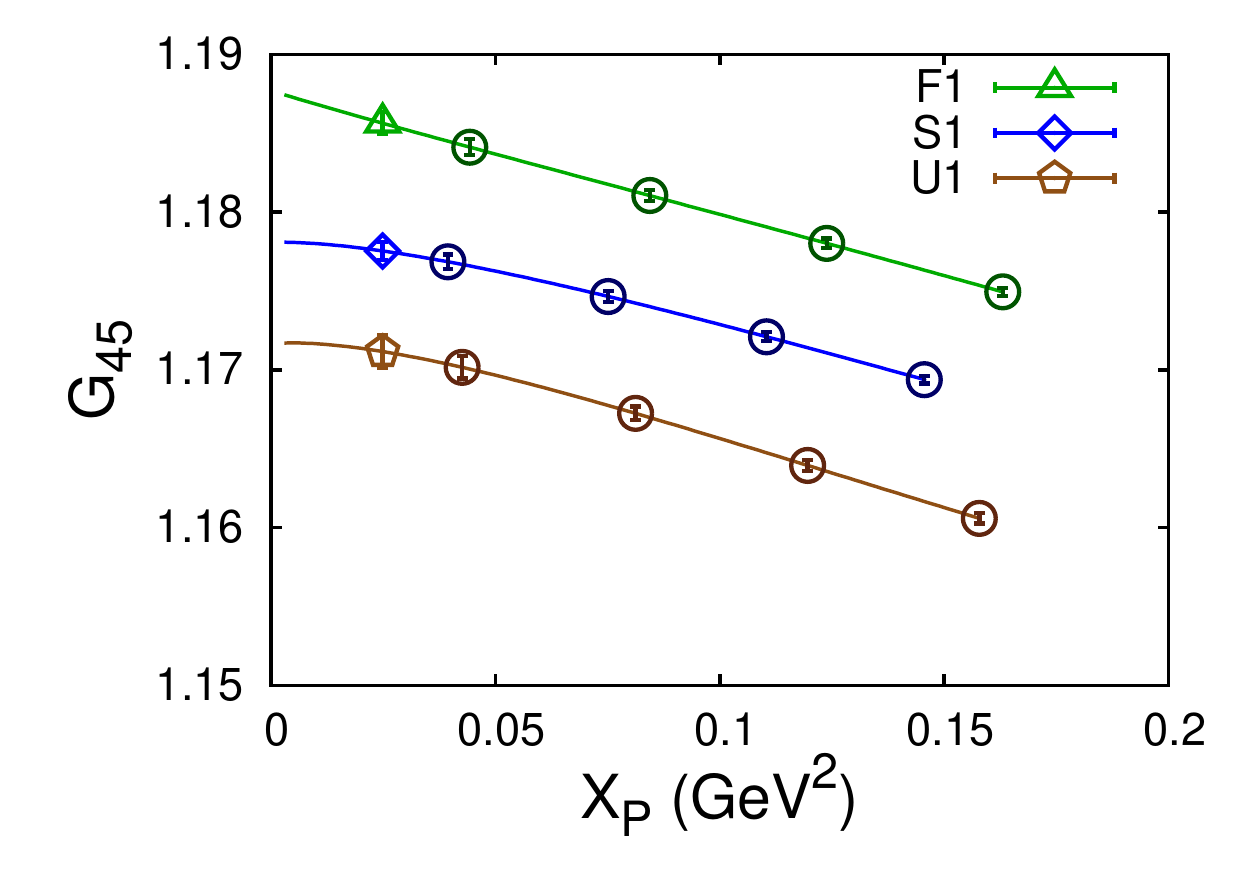}
  }
\hfill
  \subfigure[Y-fit for $G_{45}$]{
    \label{fig:G45:yfit}
    \includegraphics[width=0.48\textwidth]
                    {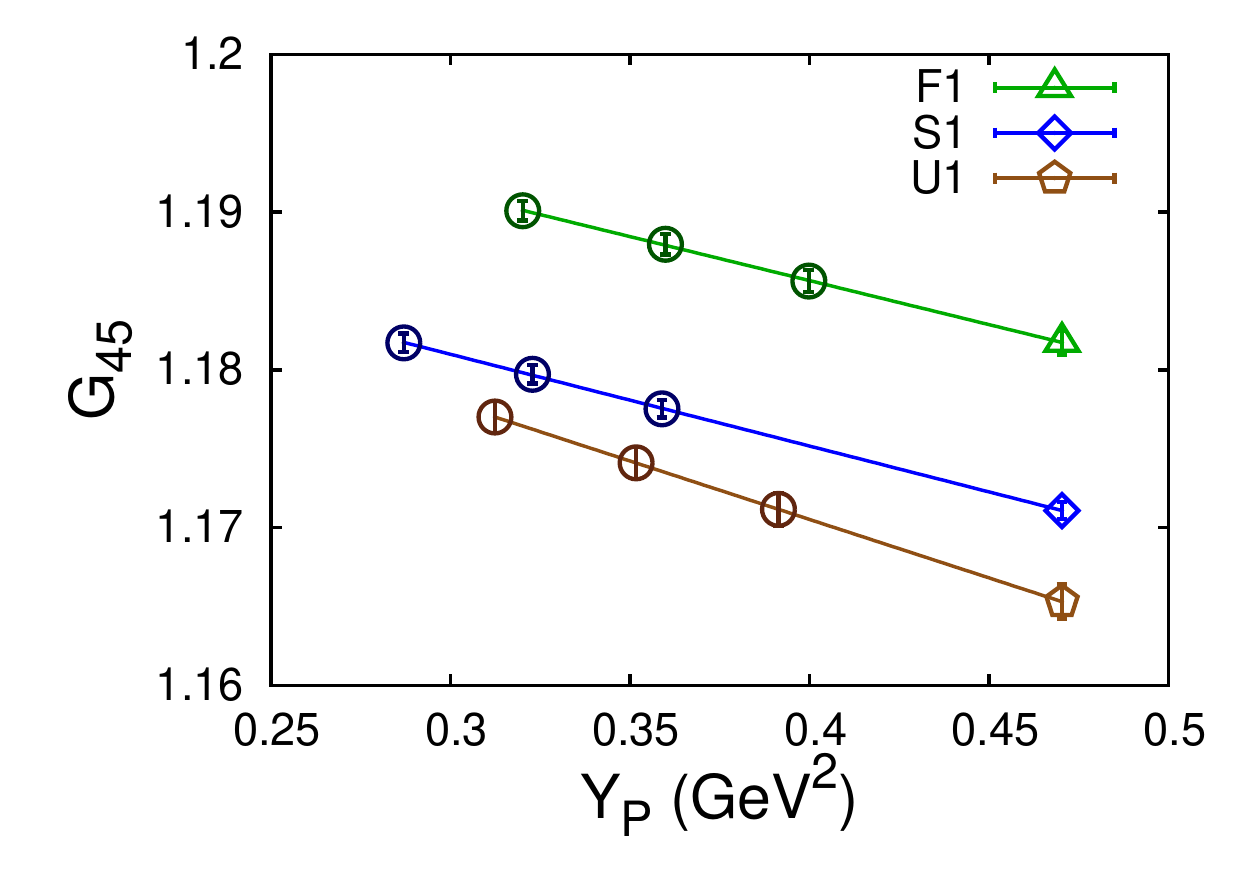}
  }
  \caption{ \subref{fig:G45:xfit} X fits and
    \subref{fig:G45:yfit} Y fits for $G_{45}$.
    Notation as in 
    Fig.~\ref{fig:BK:XYfit}.}
  \label{fig:G45:XYfit}
\end{figure*}

Examples of the X fits are shown in
Figs.~\ref{fig:BK:XYfit}\subref{fig:BK:xfit}, 
\ref{fig:G23:XYfit}\subref{fig:G23:xfit},
 and \ref{fig:G45:XYfit}\subref{fig:G45:xfit},
for $B_K$, $G_{23}$ and $G_{45}$, respectively.
\sharpe{In all these fits it was sufficient to use the narrowest
range of the Bayesian priors ($\sigma_i=1$) in order to obtain good fits.}
We note that the statistical errors appear larger in the results
for $G_{23}$ because of the finer vertical scale.
The figures emphasize the fact 
that the extrapolation in $X_P$ is relatively short.
Thus the dependence on SChPT is relatively mild,
except for the taste-breaking correction that we make to $B_K$.

In order to estimate the systematic uncertainty in the X fits
we consider two variations in the fitting scheme.
The first error is obtained from the changes in the $B_i$ and $G_j$ when
the prior widths $\sigma_a$ are doubled.
The second is obtained by repeating the fits keeping
only one NNLO term,
\begin{align}
\label{eq:X-fit-NNLO}
 B_K(\text{NNLO}) 
  &= c_1 F_0(K) + c_2 X + c_3 X^2\,, \\
 G_i(\text{NNLO}) 
  &= c_1 + c_2 X + c_3 X^2\,,
\end{align}
and using the eigenmode shift (ES) method introduced in 
Ref.~\cite{Jang:2011fp}.
The ES method tunes the fitting function in the direction of 
the eigenvectors of the covariance matrix corresponding to 
the small eigenvalues, with small shifting parameters $\eta$ 
that are constrained by the Bayesian prior condition: 
$\eta = 0 \pm \sigma_\eta$.
We set $\sigma_\eta$ from the size of the neglected highest 
order term in the fitting function, 
\begin{align}
 \sigma_\eta = 0.006 \approx X^2(\ln(X))^2,
\end{align}
where $X\approx 0.02$. 

The total systematic error from the X fits is then obtained by 
adding these two error estimates in quadrature.
The resulting errors are discussed in Sec.~\ref{sec:final}.

We next extrapolate $m_y$ to $m_s^{\rm phys}$, using the three heaviest values
of the valence quark masses. This we denote the ``Y fit".
We expect the $B_i$ and $G_j$ to be smooth, analytic 
functions of $Y_P$, since the strange quark is far from the chiral limit.
Empirically, linear fitting works very well,
as illustrated in
Figs.~\ref{fig:BK:XYfit}\subref{fig:BK:yfit}, 
\ref{fig:G23:XYfit}\subref{fig:G23:yfit} and \ref{fig:G45:XYfit}\subref{fig:G45:yfit}.
To avoid the problem of small eigenvalues, 
we use uncorrelated fitting for the Y fits.
In all cases, fits are stable and the fit parameters are consistent
across all lattices with a given nominal lattice spacing
(within the statistical uncertainties).
To estimate the systematic error in the results of the Y fits, 
we repeat the fits using a quadratic function of $Y_P$.
The changes in the final results for $B_i$ and $G_j$ 
are then taken as the systematic error. 
%
%

\section{Continuum-chiral extrapolation }
\label{sec:conti-chi}

The outputs of the extrapolations in valence masses are values for
the $B$-parameters and gold-plated combinations on each ensemble,
for continuum operators evaluated at the renormalization scale $\mu=1/a$. 
In order to compare these results and extrapolate them to the continuum
limit, and to physical sea-quark masses,
we must use renormalization group (RG) evolution to evolve to a common scale.
The standard choices for this scale in the literature are
$\mu=2\;\GeV$ and $\mu=3\;\GeV$, and we present results for both.
Since we use one-loop matching, to do the running consistently we need the continuum
two-loop anomalous dimension matrix. This has been calculated in
Ref.~\cite{Buras:2000if} for a particular choice of evanescent operators.
Because of this, it is essential that our lattice-continuum matching
uses the same set of evanescent operators, as is indeed the case in
Ref.~\cite{Kim:2014tda}.
Some technical issues arise in the RG running; these are
described in Ref.~\cite{Kim:2014tda} along with our resolutions.

We present our results for $B_K$ and the gold-plated combinations $G_i$ 
at the two renormalization scales in
Tables~\ref{tab:bk_g-2gev:all} and~\ref{tab:bk_g-3gev:all}.
Statistical errors range from the percent level to an order of magnitude smaller.
We have also obtained results for the $B_j$ ($j=2-5$) but do not show these
as they are not used in our final analysis.

%
\begin{table*}[tbhp]
\caption{ $B_K$ and gold-plated combinations for $\mu=2\GeV$ on each
  lattice listed in Table~\protect\ref{tab:milc-lat}. \sharpe{ The
    superscripts indicate whether broadened
    Bayesian priors have been used in the X-fits: 
    $\dagger$ implying $c_{4-6}=0\pm2$, 
    while $\ddagger$ implying $c_{4-6}=0\pm4$.
    Results without superscripts are obtained with $c_{4-6}=0\pm1$.}  
}
\label{tab:bk_g-2gev:all}
\begin{ruledtabular}
\begin{tabular}{ l | c c c c c }
ID		&	$B_K$						&	$G_{21}$					&	$G_{23}$			&	$G_{24}$			&	$G_{45}$			\\
\hline
C1		&	0.5484(55)					&	\ben{$0.995(11)^\dagger$}	&	1.4140(10)			&	0.6205(19)			&	\sharpe{1.1836(7)}	\\
C2		&	0.5528(56)					&	\ben{$0.993(11)^\dagger$}	&	1.4119(11)			&	0.6232(22)			&	\sharpe{1.1824(8)}	\\
C3		&	0.5673(52)					&	\ben{$0.975(10)^\ddagger$}	&	\sharpe{1.4098(9)}	&	0.6256(20)			&	\sharpe{1.1819(7)}	\\
C3-2	&	0.5715(51)					&	\sharpe{$0.974(9)^\dagger$}	&	\sharpe{1.4105(9)}	&	0.6229(16)			&	\sharpe{1.1823(6)}	\\
C4		&	0.5641(54)					&	\ben{$0.987(11)^\ddagger$}	&	\sharpe{1.4113(9)}	&	0.6291(19)			&	\sharpe{1.1822(6)}	\\
C5		&	0.5677(46)					&	\sharpe{$0.976(8)^\dagger$}	&	\sharpe{1.4082(8)}	&	0.6264(15)			&	\sharpe{1.1834(5)}	\\
\hline           
F1		&	0.5294(43)					&	1.0451(69)					&	1.4000(10)			&	0.6092(19)			&	\sharpe{1.2003(9)}	\\
F2		&	0.5451(35)					&	1.0281(59)					&	\sharpe{1.3985(6)}	&	0.6088(11)			&	\sharpe{1.1993(5)}	\\
F3		&	0.5226(49)					&	\ben{$1.053(10)^\dagger$}	&	1.4015(11)			&	0.6119(21)			&	1.1991(10)			\\
F4		&	0.5255(30)					&	\ben{$1.0366(64)^\ddagger$}	&	\sharpe{1.4033(7)}	&	0.6050(12)			&	\sharpe{1.2008(6)}	\\
F5		&	0.5388(43)					&	\ben{$1.0322(84)^\dagger$}	&	\sharpe{1.3995(9)}	&	0.6101(17)			&	\sharpe{1.1997(8)}	\\
F6		&	\ben{$0.5472(59)^\dagger$}	&	\ben{$1.014(11)^\ddagger$}	&	1.3991(13)			&	0.6123(23)			&	\sharpe{1.2018(9)}	\\
F7		&	0.5392(35)					&	1.0394(59)					&	\sharpe{1.3953(7)}	&	0.6130(12)			&	\sharpe{1.1992(6)}	\\
F9		&	0.5501(16)					&	1.0258(30)					&	\sharpe{1.3976(3)}	&	\sharpe{0.6093(6)}	&	\sharpe{1.1991(3)}	\\
\hline           
S1		&	0.5359(38)					&	1.0531(55)					&	1.4140(10)			&	0.5858(17)			&	\sharpe{1.2288(8)}	\\
S2		&	0.5361(36)					&	1.0423(57)					&	\sharpe{1.4116(9)}	&	0.5833(13)			&	\sharpe{1.2278(8)}	\\
S3		&	\ben{$0.5261(41)^\dagger$}	&	\ben{$1.0625(79)^\ddagger$}	&	1.4184(13)			&	0.5842(20)			&	1.2278(10)			\\
S4		&	0.5204(33)					&	1.0621(60)					&	1.4124(10)			&	0.5820(18)			&	\sharpe{1.2277(8)}	\\
S5		&	0.5384(36)					&	1.0446(55)					&	\sharpe{1.4110(8)}	&	0.5835(12)			&	\sharpe{1.2287(8)}	\\
\hline           
U1		&	0.5325(70)					&	\ben{1.056(11)}				&	1.4302(28)			&	0.5718(39)			&	1.2539(19)
\end{tabular}
\end{ruledtabular}
\end{table*}
%

%
\begin{table*}[tbhp]
\caption{$B_K$ and gold-plated combinations for $\mu=3\GeV$ on each
  lattice listed in Table~\protect\ref{tab:milc-lat}. \wlee{The
    convention for $\dagger$ and $\ddagger$ is the same as Table
    \protect\ref{tab:bk_g-2gev:all}. } }
\label{tab:bk_g-3gev:all}
\begin{ruledtabular}
\begin{tabular}{ l | c c c c c }
ID		&	$B_K$						&	$G_{21}$					&	$G_{23}$			&	$G_{24}$			&	$G_{45}$			\\
\hline
C1		&	0.5298(53)					&	\ben{$0.951(10)^\dagger$}	&	\sharpe{1.3942(8)}	&	0.5713(18)			&	\sharpe{1.1468(6)}	\\
C2		&	0.5341(54)					&	\ben{$0.950(10)^\dagger$}	&	\sharpe{1.3926(8)}	&	0.5738(20)			&	\sharpe{1.1459(6)}	\\
C3		&	0.5481(50)					&	\ben{$0.9323(97)^\ddagger$}	&	\sharpe{1.3911(7)}	&	0.5760(19)			&	\sharpe{1.1455(5)}	\\
C3-2	&	0.5521(49)					&	\ben{$0.9308(89)^\dagger$}	&	\sharpe{1.3915(7)}	&	0.5735(14)			&	\sharpe{1.1458(5)}	\\
C4		&	0.5449(52)					&	\ben{$0.944(10)^\ddagger$}	&	\sharpe{1.3921(7)}	&	0.5792(18)			&	\sharpe{1.1457(5)}	\\
C5		&	0.5484(45)					&	\ben{$0.9327(80)^\dagger$}	&	\sharpe{1.3898(6)}	&	0.5767(14)			&	\sharpe{1.1467(4)}	\\
\hline         
F1		&	0.5115(42)					&	0.9991(66)					&	\sharpe{1.3829(7)}	&	0.5610(18)			&	\sharpe{1.1594(7)}	\\
F2		&	0.5266(34)					&	0.9828(56)					&	\sharpe{1.3817(5)}	&	0.5606(10)			&	\sharpe{1.1586(4)}	\\
F3		&	0.5049(47)					&	\ben{$1.0069(96)^\dagger$}	&	\sharpe{1.3840(9)}	&	0.5634(19)			&	\sharpe{1.1584(8)}	\\
F4		&	0.5077(29)					&	\ben{$0.9909(61)^\ddagger$}	&	\sharpe{1.3853(5)}	&	0.5571(11)			&	\sharpe{1.1597(4)}	\\
F5		&	0.5205(42)					&	\ben{$0.9867(80)^\dagger$}	&	\sharpe{1.3825(6)}	&	0.5618(16)			&	\sharpe{1.1589(6)}	\\
F6		&	\ben{$0.5287(57)^\dagger$}	&	\ben{$0.969(10)^\ddagger$}	&	1.3821(10)			&	0.5639(21)			&	\sharpe{1.1604(7)}	\\
F7		&	0.5210(34)					&	0.9935(56)					&	\sharpe{1.3793(5)}	&	0.5645(11)			&	\sharpe{1.1585(5)}	\\
F9		&	0.5314(16)					&	0.9806(29)					&	\sharpe{1.3811(3)}	&	\sharpe{0.5611(5)}	&	\sharpe{1.1585(2)}	\\
\hline         
S1		&	0.5178(37)					&	1.0068(53)					&	\sharpe{1.3927(8)}	&	0.5394(16)			&	\sharpe{1.1806(6)}	\\
S2		&	0.5179(34)					&	0.9965(55)					&	\sharpe{1.3909(7)}	&	0.5372(12)			&	\sharpe{1.1798(6)}	\\
S3		&	\ben{$0.5083(39)^\dagger$}	&	\ben{$1.0158(76)^\ddagger$}	&	1.3960(10)			&	0.5380(19)			&	\sharpe{1.1798(8)}	\\
S4		&	0.5028(31)					&	1.0153(58)					&	\sharpe{1.3916(8)}	&	0.5359(16)			&	\sharpe{1.1797(6)}	\\
S5		&	0.5202(34)					&	0.9986(53)					&	\sharpe{1.3905(6)}	&	0.5373(12)			&	\sharpe{1.1805(6)}	\\
\hline         
U1		&	0.5145(67)					&	1.009(10)					&	1.4044(21)			&	0.5266(36)			&	1.1991(14)

\end{tabular}
\end{ruledtabular}
\end{table*}

The final step of our analysis is to do a simultaneous extrapolation
to the physical values of the sea-quark masses and to the continuum limit.
We call this procedure ``the continuum-chiral extrapolation'', although
this name is slightly misleading as the valence chiral extrapolation has already
been done.
As substitutes for sea quark masses, we use $L_P$ and $S_P$, which are,
respectively,
the squared masses of taste-$\xi_5$ (Goldstone) pions composed of two light sea
quarks ($l\bar{l}$) and two strange sea quarks ($s\bar{s}$).
They are extrapolated to their physical values, which we take to be
$m_{\pi_0}^2 = (0.1349766\GeV)^2$ for $L_P$ and
$M_{ss,\rm phys}^2 = (0.6858\GeV)^2$ for $S_P$~\cite{Davies:2009tsa}.

We expect the dependence of the $B_i$ and $G_j$ on $L_P$, $S_P$ and
$a^2$ to be analytic, with terms organized according to standard SChPT
power counting.  At NLO, the only term in SChPT that could violate
this expectation is the chiral logarithm. This is absent for the
$G_j$.  For the $B_i$, as shown by Eq.~(\ref{eq:f0}), the only
logarithms that appear have the schematic dependence $(L_P + a^2) \log
X_B$ and $X_B \log X_B$.  Since $X_B$ is set by hand to its physical
value, the $a^2$ dependence it contains is removed.  The remaining
dependence on $L_P$ and $a^2$ is analytic, and in fact also is removed
by hand when we set $L_P$ to its physical value and $a^2$ to zero.
Chiral logarithms of higher order can lead to non-analyticities, or
large derivatives, but these are numerically suppressed.  Thus, to
good approximation, we expect all the quantities we calculate to be
described by
\begin{align}
\label{eq:b_lp_sp_scaling_lin_f1}
 \tilde{F}_1 = d_1 + d_2 \frac{L_P}{\Lambda_\chi^2} 
     + d_3 \frac{S_P-M_{ss,\rm phys}^2 }{\Lambda_\chi^2} 
     + d_4(a \Lambda_Q)^2
\,.
\end{align}
Here $\Lambda_Q=0.3\;$GeV and $\Lambda_\chi=1\;$GeV,
and we have chosen to expand the $d_3$ term about the physical
$s\bar s$ mass.

When we fit our results to this form,
we impose Bayesian constraints on the linear terms
to enforce the expected power counting: $d_{2-4} = 0 \pm 2$.
\sharpe{We have also tried fits with broader contraints, $d_{2-4}=0\pm 4$,
but find that these do not significantly change the $\chi^2$ or
the resulting fit parameters.}
We find, as was the case in our earlier 
work~\cite{Bae:2011ff,Bae:2013tca,Bae:2014sja} 
that we cannot obtain a good description if we include the coarse lattices. 
Thus we fit all the fine, superfine and ultrafine lattice data to
Eq.~(\ref{eq:b_lp_sp_scaling_lin_f1}).
We call this the  $\tilde{F}_1$ fit, since it is a small variation from the fitting
function $F^1_B$ in our previous work \cite{Bae:2014sja}
(differing only in the offset in the $d_3$ term).
Since the number of configurations differ on each ensemble, errors on the
fit parameters are obtained using a variant of the bootstrap method.
Note that for this fit there are no correlations between the different ensembles.

\begin{table*}[tbhp]
\caption{Results of $\tilde{F}_1$ fits to $B_K$ and the gold-plated combinations.
 The renormalization scale is $\mu=2\GeV$.
  \label{tab:b_lp_sp_scaling_lin_f1}}
\begin{ruledtabular}
\begin{tabular}{ l | r r r r r }
                    & $B_K$     & $G_{21}$  & $G_{23}$  & $G_{24}$  & $G_{45}$  \\
\hline
$d_1$               &  \ben{0.5390(37)}&  \ben{1.0568(62)}&  \sharpe{1.4248(10)}&  \ben{0.5590(15)}&  \sharpe{1.2567(8)}\\
$d_2$               & -0.127(14)&  \ben{0.095(27)}&  \ben{0.0275(33)}& \ben{-0.0097(56)}&  \ben{0.0041(26)}\\
$d_3$               &  \ben{0.006(15)}& \ben{-0.014(25)}&  \ben{0.0145(30)}& \ben{-0.0207(53)}&  \ben{0.0026(25)}\\
$d_4$               &  \ben{ 0.78(25)}&  \ben{-1.92(41)}&  \ben{-1.799(64)}&   \ben{3.21(10)}&  \ben{-3.529(53)}\\
\hline
$B_K$ or $G_i$      & \ben{0.5366(36)}& \ben{1.0585(59)}& \sharpe{1.4253(10)}& \ben{0.5589(15)}& \sharpe{1.2568(8)}\\
\hline
$\chi^2/\text{dof}$ &       \ben{1.53}&       \ben{1.30}&       2.01&       1.08&       4.07
\end{tabular}
\end{ruledtabular}
\end{table*}
%

\begin{figure*}[tbhp]
  \renewcommand{\subfigcapskip}{-0.55em}
  \subfigure[\ben{$\tilde{F}_1$: $\chi^2/\text{d.o.f} = 1.53$}]{
    \label{fig:BK:F1:2GeV}
    \includegraphics[width=0.48\textwidth]
                    {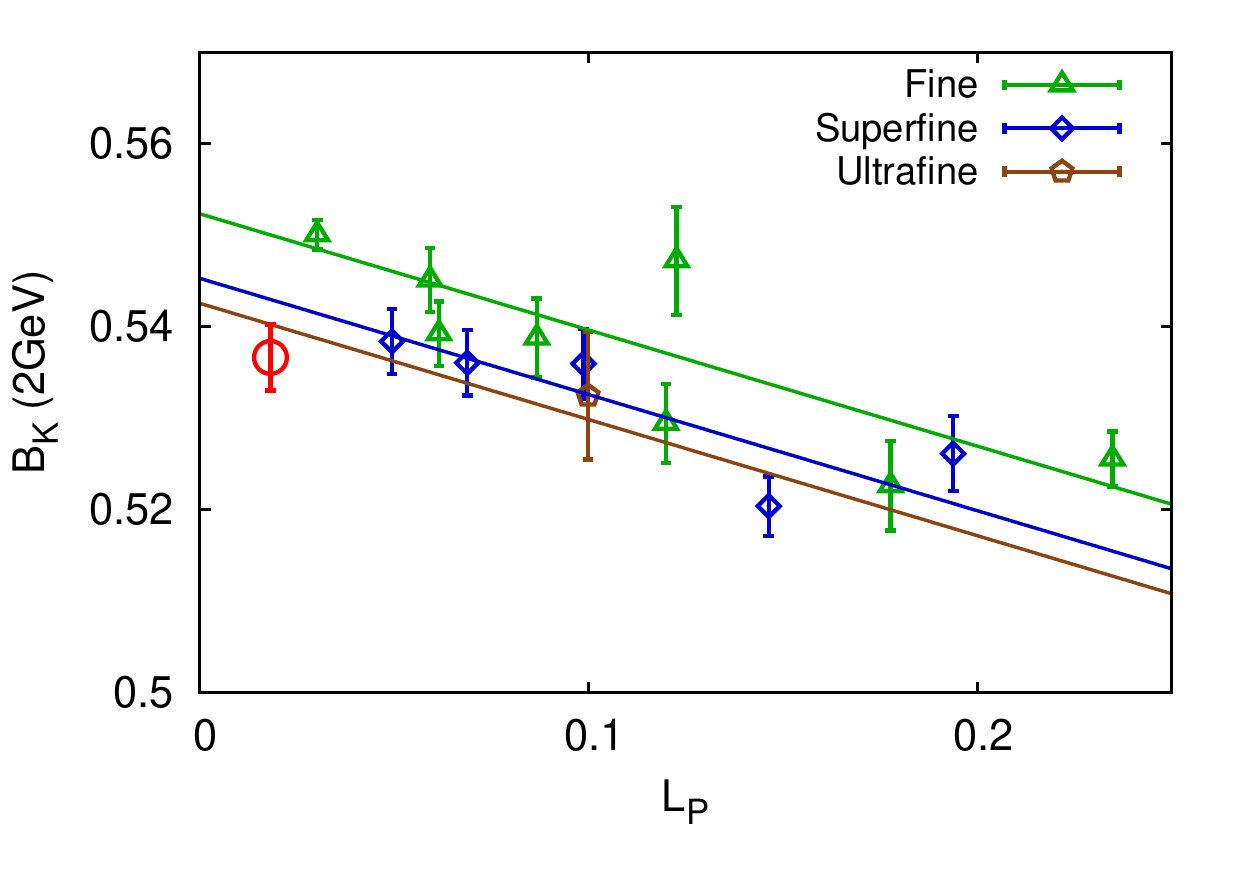}
  }
  \hfill
  \subfigure[\ben{$\tilde{F}_4$: $\chi^2/\text{d.o.f} =1.52$}]{
    \label{fig:BK:F4:2GeV}
    \includegraphics[width=0.48\textwidth]
                    {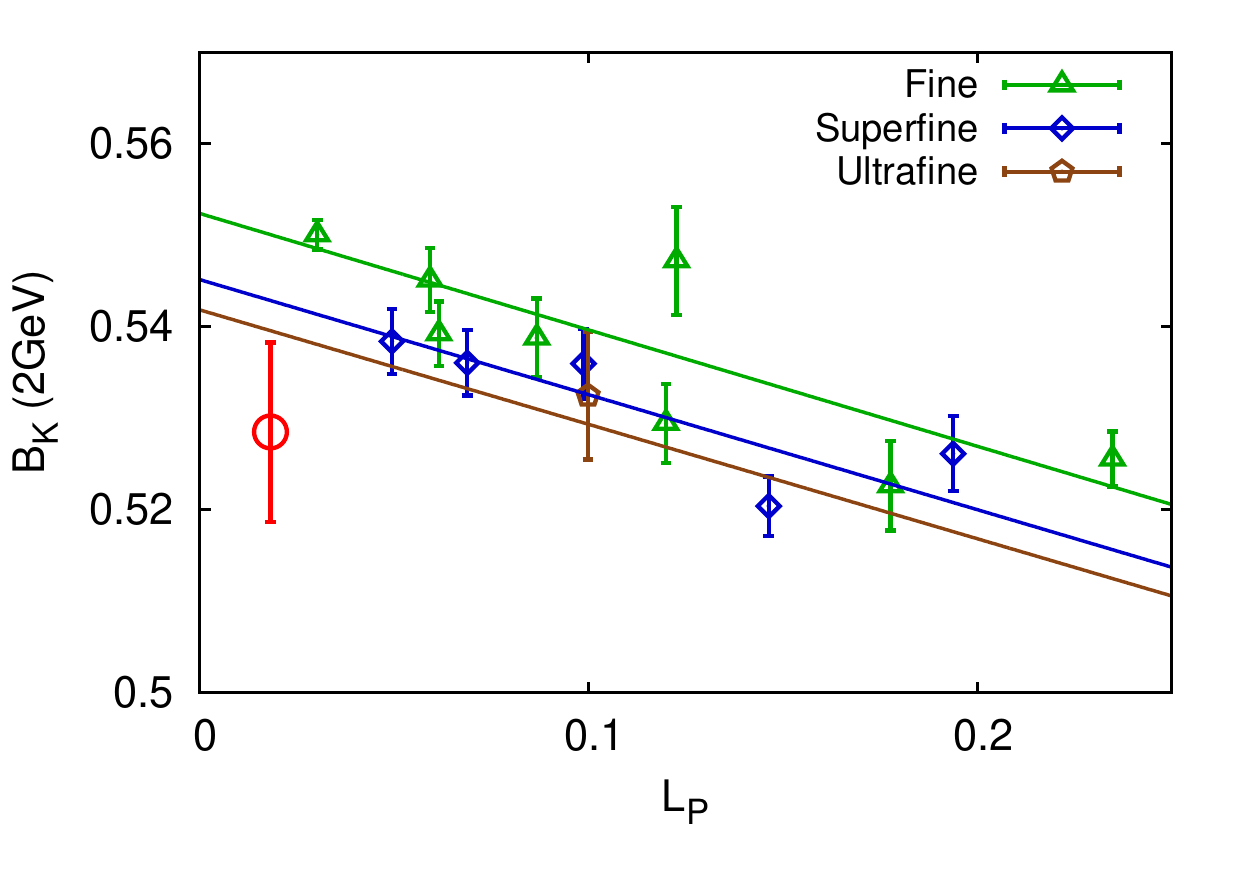}
  }
  \caption{ Continuum-chiral extrapolation for $B_K$ renormalized at
    $\mu=2\GeV$.  Results from the fine, superfine and ultrafine lattices are
    shown with (green) triangles, (blue) diamonds and the (brown)
    pentagon, respectively. (a) $\tilde{F}_1$ fit; (b) $\tilde{F}_4$ fit.
    The (red) circle gives the extrapolated result.
    Due to the variations in values of $S_P$ and $a$, the curves should not
    pass precisely through all the points. For more discussion, see text.}
  \label{fig:BK:F1+F4:2GeV}
\end{figure*}
%

\begin{figure*}[tbhp]
  \renewcommand{\subfigcapskip}{-0.55em}
  \subfigure[\ben{$\tilde{F}_1$: $\chi^2/\text{d.o.f} = 1.30$}]{
    \label{fig:G12:F1:2GeV}
    \includegraphics[width=0.48\textwidth]
                    {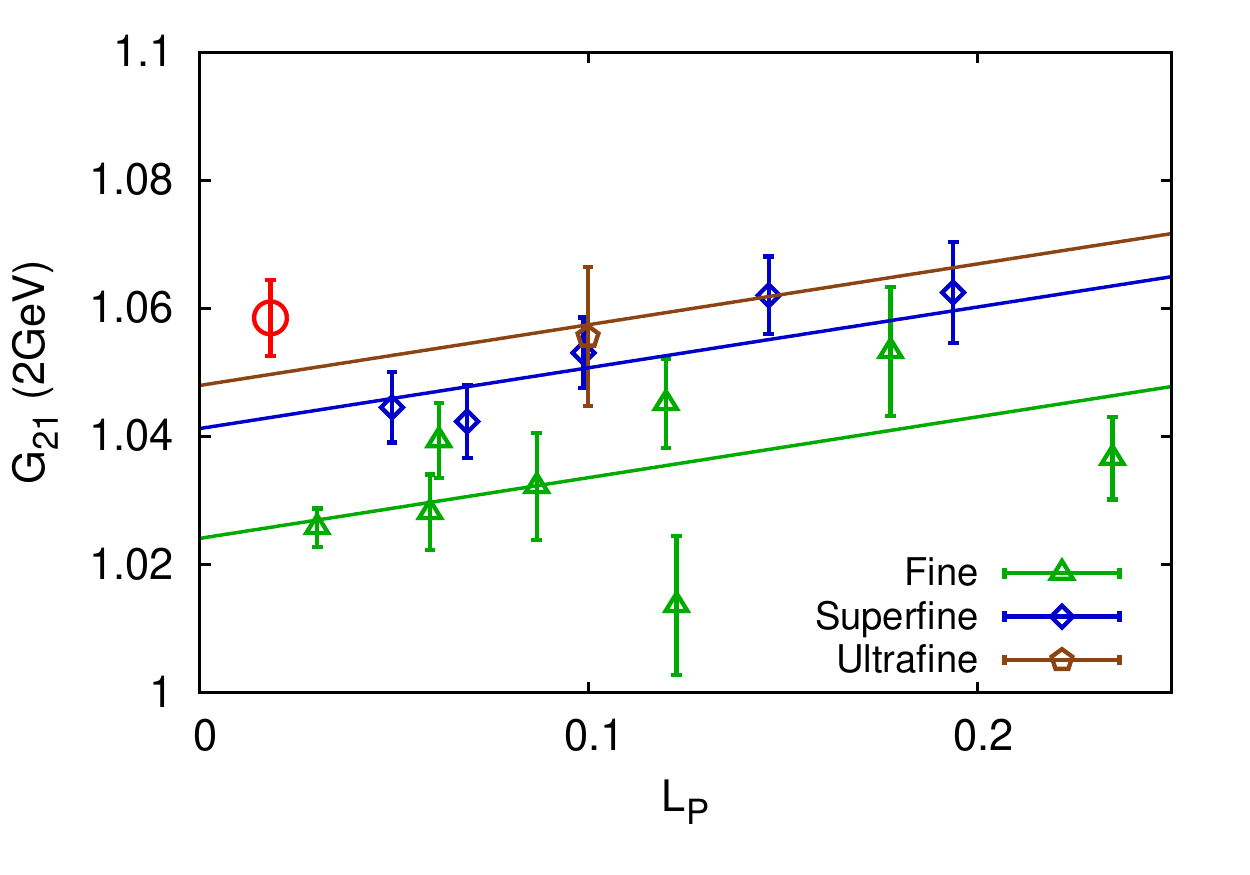}
  }
  \hfill
  \subfigure[\ben{$\tilde{F}_4$: $\chi^2/\text{d.o.f} =1.23$}]{
    \label{fig:G12:F4:2GeV}
    \includegraphics[width=0.48\textwidth]
                    {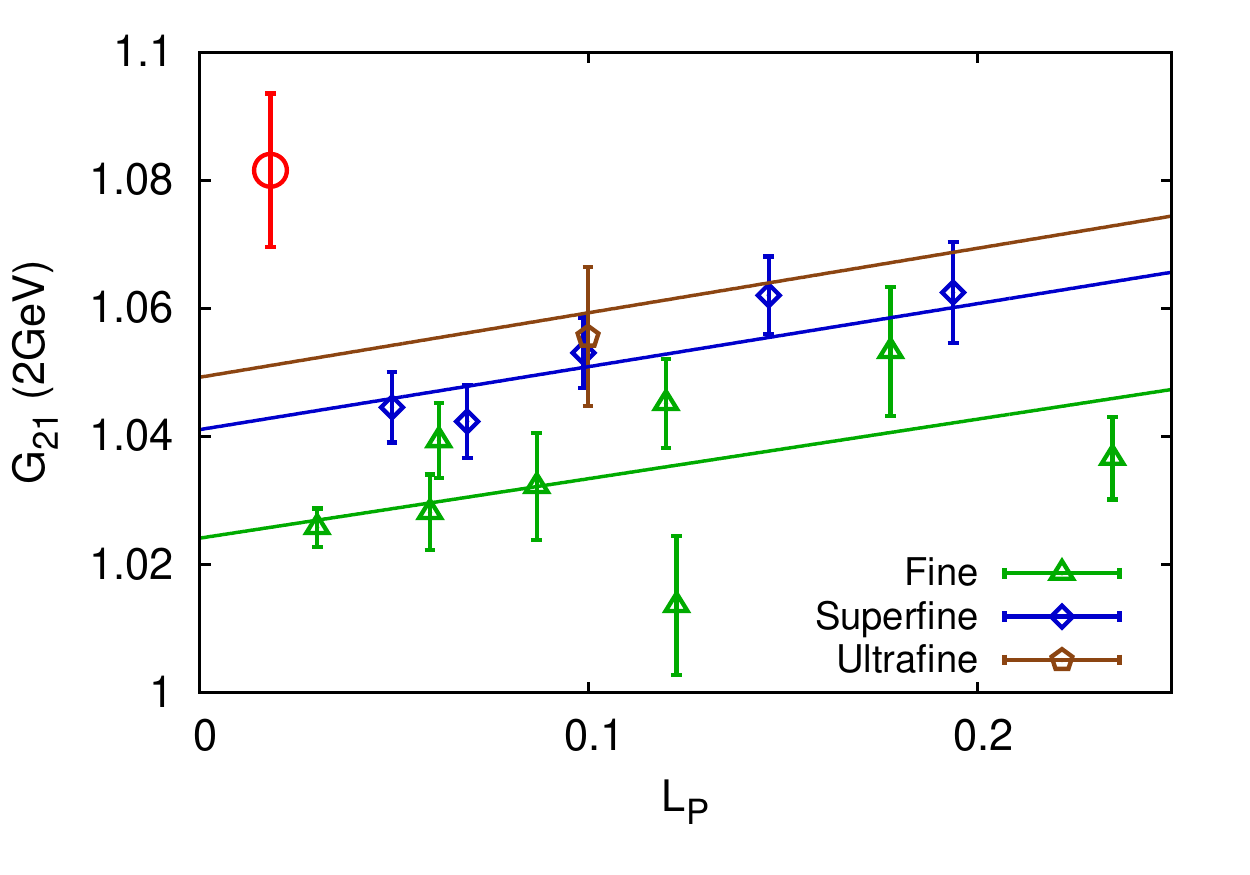}
  }
  \caption{ Continuum-chiral extrapolation results for \wlee{$G_{21}$}
    at $\mu=2\GeV$.  The notation is as in
    Fig.~\ref{fig:BK:F1+F4:2GeV}. }
  \label{fig:G12:F1+F4:2GeV}
\end{figure*}
%

\begin{figure*}[tbhp]
  \renewcommand{\subfigcapskip}{-0.55em}
  \subfigure[$\tilde{F}_1$: $\chi^2/\text{d.o.f} = 2.01$]{
    \label{fig:G23:F1:2GeV}
    \includegraphics[width=0.48\textwidth]
                    {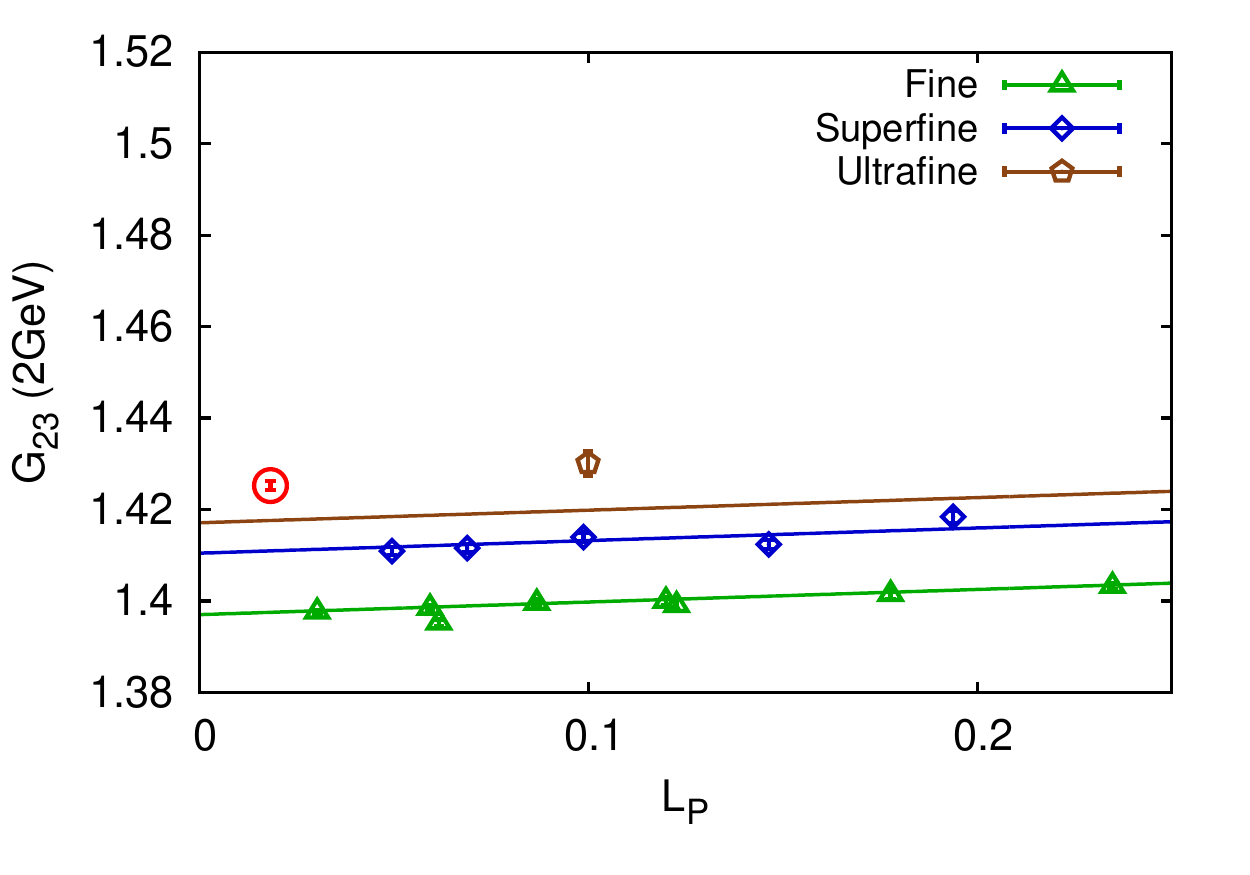}
  }
  \hfill
  \subfigure[$\tilde{F}_4$: $\chi^2/\text{d.o.f} =1.33$]{
    \label{fig:G23:F4:2GeV}
    \includegraphics[width=0.48\textwidth]
                    {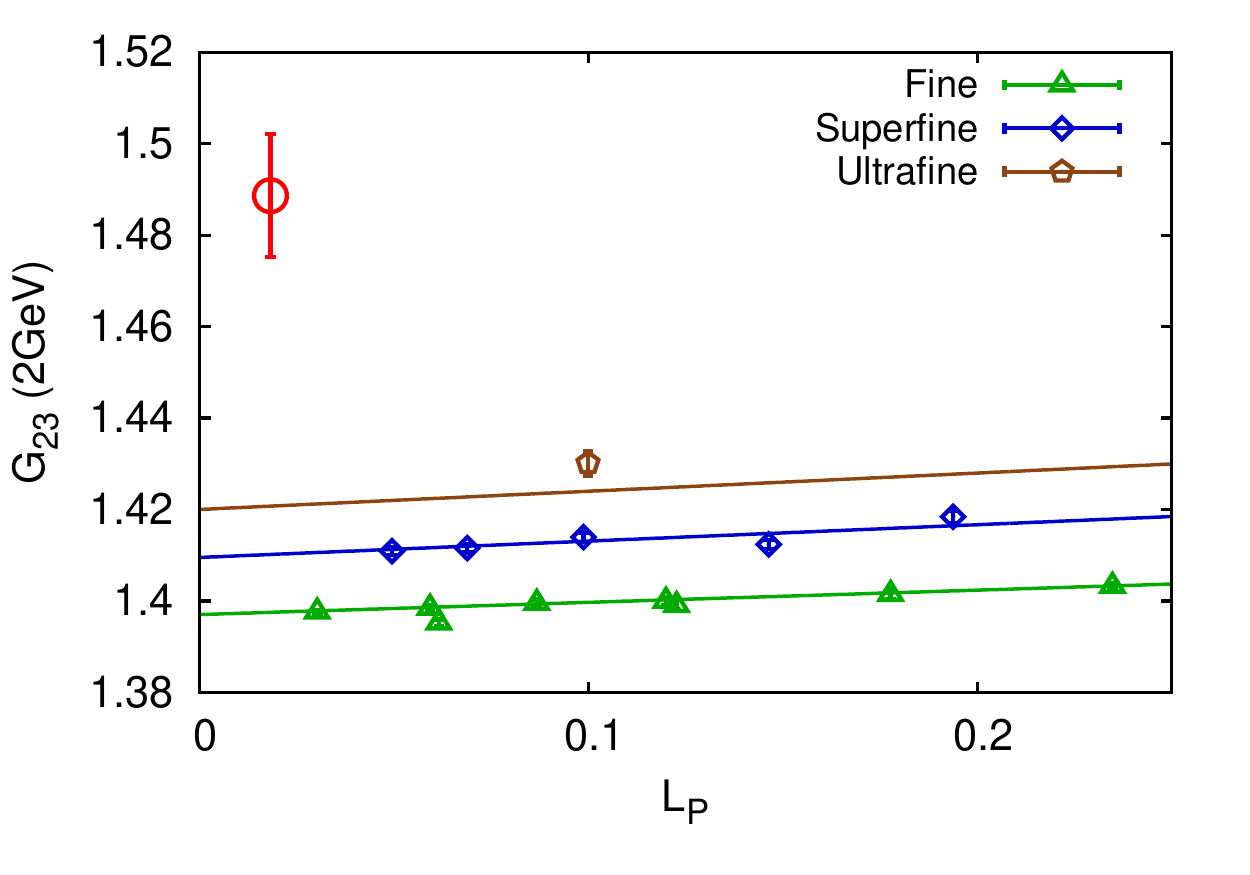}
  }
  \caption{ Continuum-chiral extrapolation results for $G_{23}$ at
    $\mu=2\GeV$.  The notation is as 
  in Fig.~\ref{fig:BK:F1+F4:2GeV}. }
  \label{fig:G23:F1+F4:2GeV}
\end{figure*}
%

\begin{figure*}[tbhp]
  \renewcommand{\subfigcapskip}{-0.55em}
  \subfigure[$\tilde{F}_1$: $\chi^2/\text{d.o.f} = 1.08$]{
    \label{fig:G24:F1:2GeV}
    \includegraphics[width=0.48\textwidth]
                    {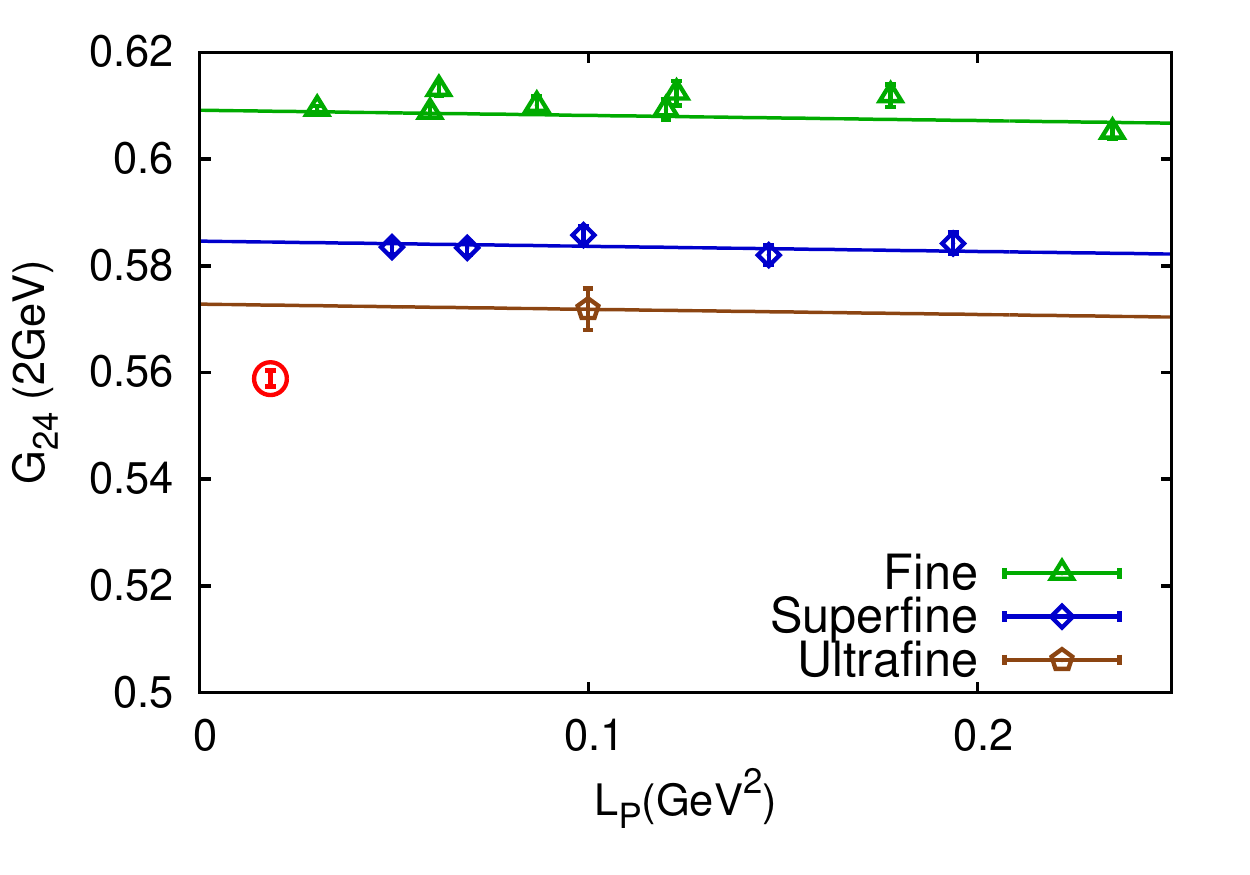}
  }
  \hfill
  \subfigure[$\tilde{F}_4$: $\chi^2/\text{d.o.f} =0.91$]{
    \label{fig:G24:F4:2GeV}
    \includegraphics[width=0.48\textwidth]
                    {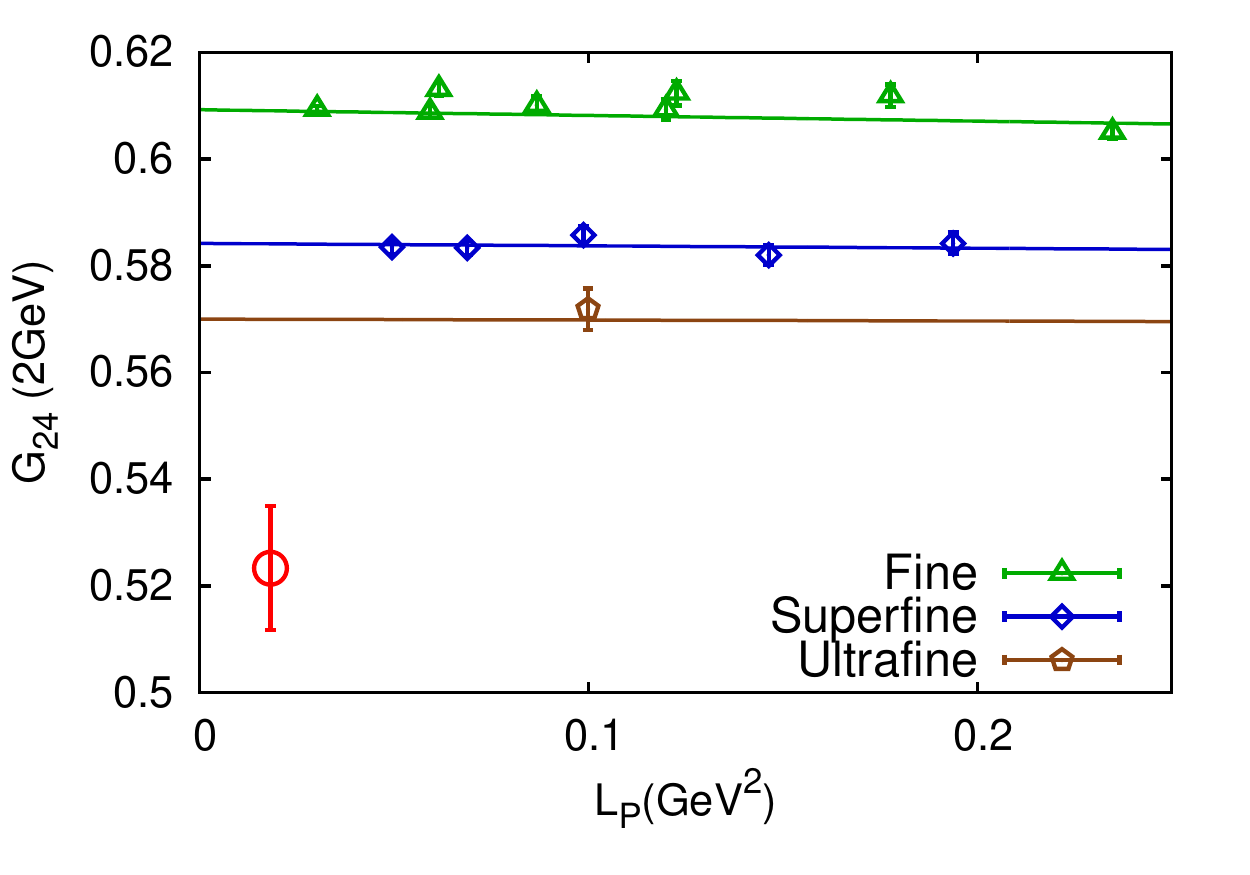}
  }
  \caption{ Continuum-chiral extrapolation results for $G_{24}$ at
    $\mu=2\GeV$.  The notation is as 
  in Fig.~\ref{fig:BK:F1+F4:2GeV}. }
  \label{fig:G24:F1+F4:2GeV}
\end{figure*}
%

\begin{figure*}[t!]
  \renewcommand{\subfigcapskip}{-0.55em}
  \subfigure[$\tilde{F}_1$: $\chi^2/\text{d.o.f} = 4.07$]{
    \label{fig:G45:F1:2GeV}
    \includegraphics[width=0.48\textwidth]
                    {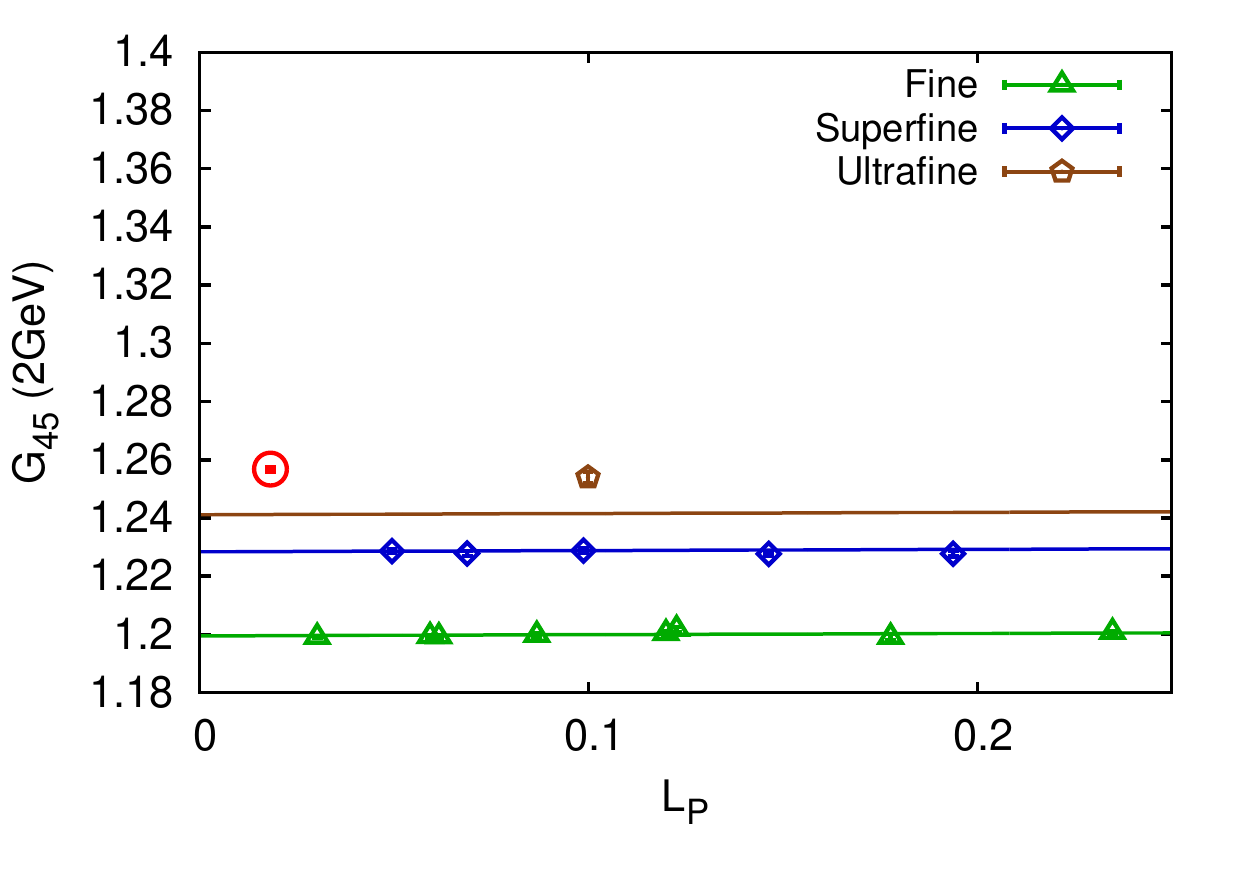}
  }
  \hfill
  \subfigure[$\tilde{F}_4$: $\chi^2/\text{d.o.f} =1.39$]{
    \label{fig:G45:F4:2GeV}
    \includegraphics[width=0.48\textwidth]
                    {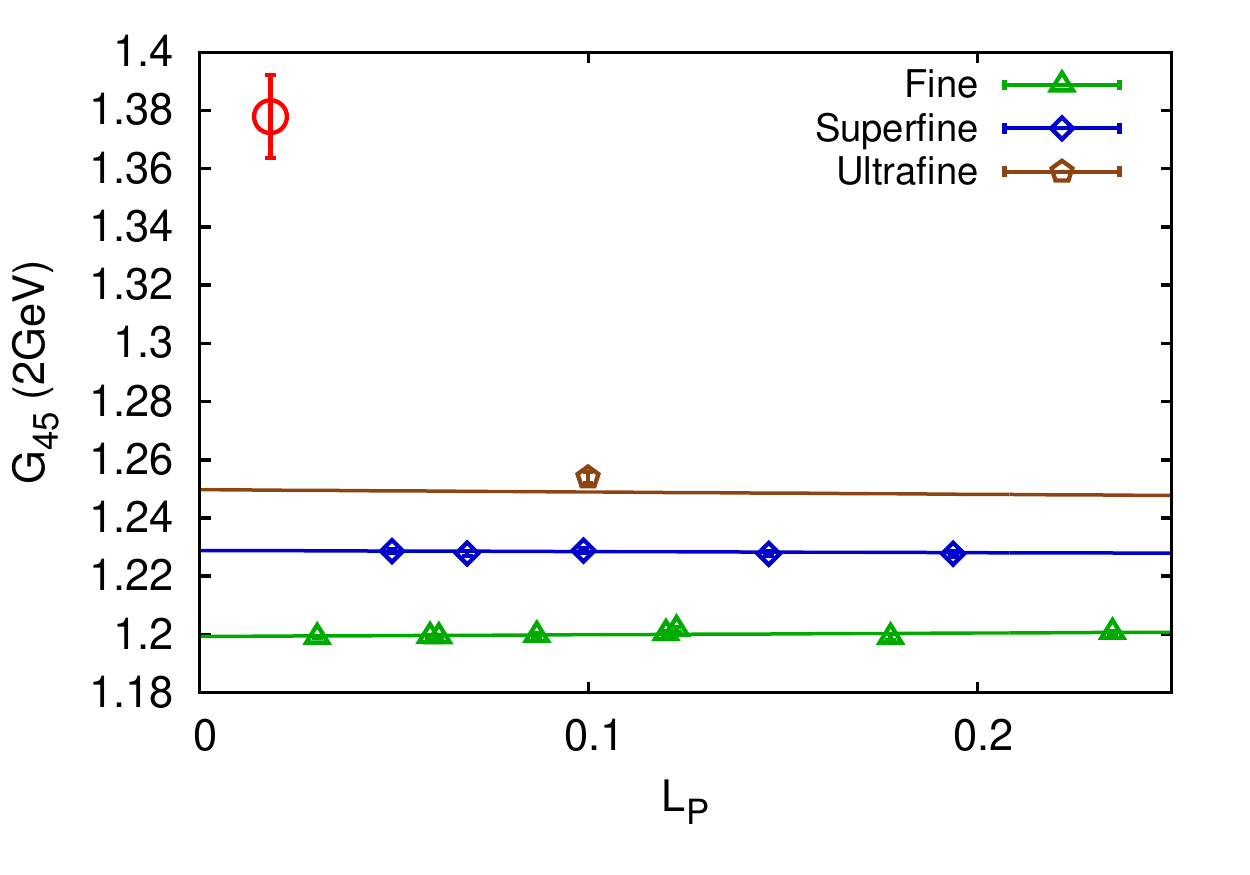}
  }
  \caption{ Continuum-chiral extrapolation results for $G_{45}$ at
    $\mu=2\GeV$.  The notation is as 
  in Fig.~\ref{fig:BK:F1+F4:2GeV}. }
  \label{fig:G45:F1+F4:2GeV}
\end{figure*}

In Table~\ref{tab:b_lp_sp_scaling_lin_f1}, we show the results of the
$\tilde{F}_1$ fits to $B_K$ and the $G_i$ (renormalized at $\mu=2\GeV$).
Plots of the fits are shown in
Figs.~\ref{fig:BK:F1+F4:2GeV}\subref{fig:BK:F1:2GeV},
\ref{fig:G12:F1+F4:2GeV}\subref{fig:G12:F1:2GeV},
\ref{fig:G23:F1+F4:2GeV}\subref{fig:G23:F1:2GeV},
\ref{fig:G24:F1+F4:2GeV}\subref{fig:G24:F1:2GeV}, and
\ref{fig:G45:F1+F4:2GeV}\subref{fig:G45:F1:2GeV}.
The fits are qualitatively similar and of comparable quality 
if the operators are renormalized at $\mu=3 \GeV$.
To interpret these plots the following must be kept in mind.
For each nominal value of $a$ (e.g. for the fine lattices)
there is a variation in the actual values of $a$ and in the values of
$S_P$. This is most significant for the ensembles F6 and F7, which
have a substantially lower strange quark mass than the other fine ensembles.
These variations are accounted for in the fit (with F6 and F7 providing
a significant lever arm to determine $d_3$), but {\em do not show up
in these two-dimensional plots}.
Indeed, the points from F6 and F7 are not included in the plots,
while the fit lines for the fine and superfine ensembles 
are shown with $a$ and $S_P$ set to their average values 
(excluding ensembles F6 and F7 for the fine lattices).
Thus, even if the fit were perfect, the fit lines would not pass through any of the points,
except for the ultrafine case.
Because of this, the fits appear slightly worse than they actually are;
the real indicator of goodness of each fit is the quoted value of $\chi^2/\text{d.o.f.}$.

The fits indicate that the dependence on the strange sea-quark mass is
very weak for all five quantities, with $|d_3|\ll 1$.
For the gold-plated combinations, the dependence on the light sea-quark mass is also weak,
much weaker within our range of parameters than the dependence on $a$.
Only for $B_K$ does the variation with $L_P$ have a similar magnitude to that with $a$.
The values of the $a^2$ coefficient, $d_4$, indicate that scale describing $a^2$ effects
ranges from $\sim 0.3\;$GeV ($|d_4|\sim 1$) up to $\sim 0.55\;$GeV ($|d_4|\sim 3.5$).
These scales are not unusual for discretization errors with improved staggered fermions.
The $\chi^2/\text{d.o.f.}$ of these fits is 
reasonable for $B_K$, $G_{21}$, and $G_{24}$.\footnote{%
\sharpe{Here we consider a value up to $\sim 1.5$ to be reasonable,
due to residual correlations between configurations.
We work with a bin size of 5, and Fig.~\ref{fig:autocorr}\ shows that
this can lead to an underestimate of the error by $\sim 25\%$
on some configurations. Consequently the $\chi^2$ will be overestimated
by $\sim 1.25^2$.}
}
Hence, we choose the results from the $\tilde{F}_1$ fits 
for our central values for these quantities.
However, we cannot use the $\tilde{F}_1$ results for $G_{23}$ and
$G_{45}$, since the fit quality is too poor.
This is primarily due to the difficulty that the fits have in reproducing the dependence on
$a$.

\begin{table*}[tbhp]
\caption{Fit results for BSM B-parameters and the gold-plated combinations 
  obtained using $\tilde{F}_4$-fit at $\mu=2\GeV$.
  \label{tab:b_lp_sp_scaling_lin_f4}}
\begin{ruledtabular}
\begin{tabular}{ l | r r r r r }
                    & $B_K$     & $G_{21}$  & $G_{23}$  & $G_{24}$  & $G_{45}$  \\
\hline                                                                   
$d_1$              &  \ben{0.5308(99)} & \ben{1.080(12)} &  1.488(14) &  0.523(12) &  1.378(14)  \\ 
$d_2$              & -0.124(18)       & \ben{0.104(28)} &  0.045(14) &  0.001(16) & -0.013(12)  \\ 
$d_3$              &  \ben{0.005(15)} & \ben{-0.011(25)} &  \sharpe{0.019(6)} & \sharpe{-0.021(6)} &  \sharpe{0.012(7)}  \\ 
$d_4$              &   \ben{0.24(40)} & \ben{-0.33(27)} &   2.22(75) &   0.84(63) &   3.70(80)  \\ 
$d_5$              &  \ben{-0.18(77)} & \ben{-0.66(48)} &  -1.10(87) &  \ben{-0.7(10)}&   1.16(78)  \\ 
$d_6$              &   0.10(10) &   \ben{-0.081(63)} &  -0.29(34) &   0.03(22) &  -0.64(42)  \\ 
$d_7$              &   0.09(21) &   \ben{-0.10(14)} &   1.20(40) &   0.33(33) &   2.06(42)  \\ 
$d_8$              &   \ben{0.22(21)} &   \ben{-0.65(20)} &  -1.80(37) &   0.98(31) &  -3.34(39)  \\ 
$d_9$              &  0.008(31) &  -0.008(21) &  0.183(60) &  0.036(50) &  0.322(64)  \\ 
\hline                                                                                 
$B_K$ or $G_i$     & \ben{0.5285(98)} & \ben{1.082(12)} & \ben{1.489(13)} & \ben{0.523(12)}& \ben{1.378(14)} \\ 
\hline                                                                                 
$\chi^2/\text{dof}$&       \ben{1.52} &        \ben{1.23} &       1.33 &       0.91 &        1.39 \\ 
\end{tabular}
\end{ruledtabular}
\end{table*}

To obtain reasonable fits for $G_{23}$ and $G_{45}$, we add
higher order terms to the fitting function, denoting the new form
$\tilde{F}_4$:
\begin{align}
\label{eq:b_lp_sp_scaling_lin_f4}
\tilde{F}_4 = \tilde{F}_1 &+ d_5(a \Lambda_Q)^2\frac{L_P}{\Lambda^2_\chi}
          + d_6(a \Lambda_Q)^2[\frac{S_P - m_{s\bar{s}}^2}{\Lambda^2_\chi}] 
\nonumber \\
          &+ d_7 (a \Lambda_Q)^2 \alpha_s 
          + d_8 \alpha_s^2 +d_9 (a \Lambda_Q)^4
\end{align}
where $\alpha_s = \alpha_s^{\overline{\text{MS}}}(1/a)$.
In other words, we add a subset of the analytic terms quadratic in
$L_P$, $S_P$ and $a^2$, as well as two terms that include logarithmic
dependence on $a$.
The $d_7$ term would be the dominant source of $a$ dependence were the
action and operators tree-level ${\cal O}(a^2)$ improved. In fact, our
valence fermion action and operators are not tree-level improved, so
we must include the pure $a^2$ $d_4$ term as well. Nevertheless, we
expect the tree-level contributions proportional to $a^2$ alone to be
small, due to the use of HYP-smeared gauge \wlee{fields.}
The $d_8$ term arises because our lattice operators are only matched to the
continuum operators at one-loop order, leaving a two-loop residual discrepancy.
In the $\tilde{F}_4$ fits we constrain $d_{2-9}$ using the Bayesian method,
choosing the prior conditions $d_{2-9}=0\pm2$.
\sharpe{Again we find that broadening the
priors does not significantly improve the fits.}

The results for the $\tilde{F}_4$ fits are shown (for $\mu=2\GeV$)
in Table~\ref{tab:b_lp_sp_scaling_lin_f4} and 
Figs.~\ref{fig:BK:F1+F4:2GeV}\subref{fig:BK:F4:2GeV},
\ref{fig:G12:F1+F4:2GeV}\subref{fig:G12:F4:2GeV},
\ref{fig:G23:F1+F4:2GeV}\subref{fig:G23:F4:2GeV},
\ref{fig:G24:F1+F4:2GeV}\subref{fig:G24:F4:2GeV}, and
\ref{fig:G45:F1+F4:2GeV}\subref{fig:G45:F4:2GeV}.
With the additional terms, we obtain reasonable values of
$\chi^2/\text{d.o.f}$ for $G_{23}$ and $G_{45}$,
and we take the resulting extrapolated values as our final results
for these two quantities.
For the other quantities, the fit quality is only slightly improved.

As is apparent, particularly from 
Figs.~\ref{fig:G23:F1+F4:2GeV}\subref{fig:G23:F4:2GeV},
\ref{fig:G24:F1+F4:2GeV}\subref{fig:G24:F4:2GeV} and
\ref{fig:G45:F1+F4:2GeV}\subref{fig:G45:F4:2GeV},
the change from $\tilde{F}_1$ to $\tilde{F}_4$ fits
has a very significant impact on the continuum extrapolation.
This is primarily due to the $d_8 \alpha_S^2$ term, which
has a rapid dependence on $a$ as $a\to 0$. We note that
the coefficients of this term in the fits to $G_{23}$ and $G_{45}$
are relatively large [although still of ${\cal O}(1)$], and this is
what leads to the large change in the extrapolated value between
the fits. We do not find the $\tilde{F}_4$ fits to provide a convincing
description of the $a$ dependence, particularly
as they depend very strongly on the result from the single
ultrafine lattice. However, we think that the conservative choice is
to use the better fit for the central value, and then to take the difference
between the two fits as an estimate of the systematic error in the
continuum-chiral extrapolation. The final results
from the two fits, and the resulting estimate of the systematic error,
are collected in Table~\ref{tab:lp_sp_scaling_diff_2GeV}.
For $G_{23}$, $G_{24}$ and $G_{45}$ this source of error dominates all
others, as discussed in the following section.

We have also used fit functions with additional higher-order terms.
These do lead to mild reductions in the values of $\chi^2/\text{d.o.f}$,
but do not lead to significant changes in the central values compared to the
$\tilde{F}_4$ fits. Thus they do not significantly change our estimates of
systematic errors. For the sake of brevity, we do not display the results of
these more elaborate fits.

\begin{table}[htbp]
\caption{Results for $B_K$ and $G_i$ 
(renormalized at $\mu=2\GeV$) from continuum-chiral extrapolation
using the $\tilde{F}_1$ and $\tilde{F}_4$ fits.
Our choices for the final central values are (red and) underlined.
$\Delta$ is the fractional systematic error in the continuum-chiral extrapolation,
and is obtained from the difference between the two fits.
\label{tab:lp_sp_scaling_diff_2GeV}}
\begin{ruledtabular}
\begin{tabular}{ l | c c c }

         & $\tilde{F}_1$ & $\tilde{F}_4$ & $\Delta(\%)$ \\
\hline
$B_K$   & \textcolor{red}{\underline{\ben{0.5366(36)}}}\com{} & \ben{0.5285(98)}\com{}  & \ben{1.52}\com{}\\
$G_{21}$ & \textcolor{red}{\underline{\ben{1.0585(59)}}}\com{} & \ben{1.082(12)}\com{} & \ben{2.18}\com{}\\
$G_{23}$ & \sharpe{1.4253(10)}\com{} & \textcolor{red}{\underline{\ben{1.489(13)}}}\com{} & \ben{4.26}\com{}\\
$G_{24}$ & \textcolor{red}{\underline{0.5589(15)}} & \ben{0.523(12)}\com{} & \ben{6.36}\com{}\\
$G_{45}$ & \sharpe{1.2568(8)}\com{} & \textcolor{red}{\underline{\ben{1.378(14)}}}\com{} & \ben{8.79}\com{}
\end{tabular}
\end{ruledtabular}
\end{table}
%

We close this section by returning to the option of directly fitting the
BSM $B$-parameters rather than using the gold-plated combinations.
In all cases we find that direct continuum-chiral fits have values of
$\chi^2/\text{d.o.f.}$ in the range $3-5$, both for $\tilde{F}_1$ and
$\tilde{F}_4$ (and more elaborate) fits.
We do not fully understand this failure of the continuum-chiral fits,
but suspect that it is related to errors in valence chiral extrapolation (X fits).
The X fits are better controlled with the gold-plated combinations.

\section{Final Results and Error Budget}
\label{sec:final}

In this section we discuss all sources of error, and give our final
results for the BSM $B$-parameters with their error budget.
Because we obtain $B^G_{2-5}$ using Eq.~\eqref{eq:b-from-gold}, we
estimate the errors in $B_K$ and the $G_i$ first, and then propagate the
errors to $B^G_{2-5}$.
Our final results for the two standard renormalization scales are given
in Tables~\ref{tab:results} and \ref{tab:results-gold},
while the final error budget is given in Table~\ref{tab:err-budget}.~\footnote{%
The result quoted here for $B_K(2\GeV)$ is obtained by 
a very slightly different analysis method than that we used previously
in Ref.~\cite{Bae:2014sja}. Thus the results differ slightly,
although they agree within the (small) statistical errors, and have almost
exactly the same total error.}

%
%
\begin{table}[tbhp]
\caption{Final results for $B_K$ and the BSM $B$-parameters 
at renormalization scales $\mu=2\GeV$  and $\mu=3\GeV$.
The first error is statistical, the second systematic.
\label{tab:results}}
\begin{ruledtabular}
\begin{tabular}{ l | c c c c }
      & $\mu=2\GeV$ & $\mu=3\GeV$ \\
\hline
$B_K$ & 0.537(4)(26)   & \ben{0.519(4)(26)}  \\
$B^G_2$ & 0.568(1)(25) & 0.525(1)(23)  \\
$B^G_3$ & \ben{0.382(4)(17)} & \ben{0.360(4)(16)}  \\
$B^G_4$ & \ben{0.984(3)(64)} & \ben{0.981(3)(62)}  \\
$B^G_5$ & \ben{0.714(7)(71)} & \ben{0.751(7)(68)}  \\
\end{tabular}
\end{ruledtabular}
\end{table}
\begin{table}[tbhp]
  \caption{Final results for the gold-plated combinations $G_i$.
  Notation as in Table~\ref{tab:results}.
  }
  \label{tab:results-gold}
\begin{ruledtabular}
\begin{tabular}{ c | l l }
      & $\mu=2\GeV$ & $\mu=3\GeV$ \\
\hline
$G_{21}$ & \ben{1.059(6)(52)}  &\ben{1.012(6)(50)} \\
$G_{23}$ & 1.489(13)(66) &1.460(14)(65)\\
$G_{24}$ & 0.559(1)(36)  &0.515(1)(32) \\
$G_{45}$ & 1.378(14)(123) &1.307(14)(107)\\
\end{tabular}
\end{ruledtabular}
\end{table}
%

\begin{table*}[htbp]
\caption{ Error budget for the $B_i$ and $G_j$ evaluated at 
renormalization scale $\mu =2 \GeV$. All entries are in percent.
}
\label{tab:err-budget}
\begin{ruledtabular}
\begin{tabular}{ l | c c c c c | c c c c | c | l }
  cause              &$B_K$  &$G_{21}$ &$G_{23}$ &$G_{24}$ &$G_{45}$ & $B^G_2$ & $B^G_3$ & $B^G_4$ & $B^G_5$ & \ben{$B_3^{G,\text{SUSY}}$} & method \\
  \hline                                                                                            
  statistics         &0.67   &   \ben{0.56}  &   \ben{0.87}  &   0.27  &  \ben{1.02}   & 0.25  & \ben{1.00}  & 0.27  & \ben{0.98} & \ben{0.66} & see text \\
$\left\{ \begin{array}{l} \text{matching} \\ \text{cont-extrap.} \end{array} \right\}$                                                                                
                     &4.40   &4.40     &4.40     &6.36     &8.79     & 4.40  & 4.45  & 6.36  & 9.63 & \ben{4.40} & ($\tilde{F}_1$ vs. $\tilde{F}_4$) or $\alpha_s^2$ (U1) \\                           
  finite volume      &\ben{0.73}   &\ben{0.17}     &0.05     &0.43     &\ben{0.04}     & \ben{0.56}  & \ben{0.52}  & \ben{0.99}  & \ben{1.02} & \ben{0.60} &(C3) vs. (C3-2) \\                                      
  X-fits             &0.05   &0.40     &0.45     &0.02     &0.96     & 0.36  & \ben{0.34}  & 0.37  & 1.23 & \ben{0.60} & change Bayes. prior \& fit method \\                                  %
  Y-fits             &2.07   &\ben{2.11}     &0.32     &0.48     &\ben{1.12}     & \ben{0.00}  & \ben{0.32}  & \ben{0.48}  & \ben{1.59} & \ben{0.22} & linear vs. quad. \\                                         
  $f_\pi$            &0.10   &0        &0        &0        &0        & 0.10  & 0.10  & 0.10  & 0.10 & \ben{0.10} & $132\;$MeV vs. $124.2\;$MeV. \\                                       
  $r_1$              &0.35   &\ben{0.28}     &0.11     &0.16     &0.21     & 0.07  & \ben{0.17}  & \ben{0.09}  & \ben{0.30} & \ben{0.01} & errors due to $r_1$ ambiguity.  \\                                
\hline                                                                                              
Total                &\ben{4.93}   &\ben{4.91}     &\ben{4.44}     &\ben{6.39}     &\ben{8.92}     & \ben{4.45}  & \ben{4.51}  & \ben{6.47}  & 9.90 & \ben{4.49} &
\end{tabular}
\end{ruledtabular}
\end{table*}

As can be seen from Table~\ref{tab:err-budget},
the statistical errors in $B_K$ and the $G_i$ are small,
ranging from $\sim 0.25\%$ to $\sim 1\%$.
The largest are those in $G_{23}$ and $G_{45}$, 
resulting from the use of the $\tilde{F}_4$ fits
for the continuum-chiral extrapolation.
We propagate the statistical errors into the $B^G_j$ using
the bootstrap method.
The larger errors in $G_{23}$ and $G_{45}$ then lead to
$B^G_3$ and $B^G_5$ having the largest statistical errors
of the BSM $B$-parameters.
In all cases, however, the statistical errors are
much smaller than those from systematic effects.

We now run through the systematic errors in the order listed
in Table~\ref{tab:err-budget}.
The dominant error is that due to the combined effect of
using one-loop matching and the continuum-chiral extrapolation.
We combine these because the $\tilde{F}_4$ fit includes the
$\alpha_s^2$ error that results from perturbative truncation,
and indeed this is the dominant contribution to the systematic
error estimate, as discussed above.
However, one can also estimate the truncation error directly,
following Ref.~\cite{Bae:2011ff},
by the size of a typical two-loop contribution:
\begin{align}
  \Delta B_i \approx B_i \times \alpha_s^2\,.
\end{align}
Here we use $\alpha_s$ in the $\MSbar$ scheme evaluated
at a scale $1/a_{\rm min}$, where $a_{\rm min}$ is our smallest lattice spacing.
This leads to a 4.4\% relative error.
To avoid double-counting, we take the larger of 
(a) the direct estimate of two-loop effects (4.4\%) and
(b) the difference between $\tilde{F}_1$ and $\tilde{F}_4$ fits.
In essence this method is using the $\tilde{F}_4$ fit to give an estimate
of the uncertainty in the coefficient of the $\alpha_s^2$ term, 
except that we do not allow this uncertainty to drop below unity. 

The above description applies to quantities we calculate directly,
namely $B_K$ and the $G_i$.
For the derived quantities $B^G_j$,
defined in Eq.~\eqref{eq:b-from-gold}, we proceed as follows.
We vary the fit choices (for the continuum-chiral extrapolation)
for each of the components of the $B^G_j$ independently,
and take the largest variation from the central value as the
systematic error estimate.
If this maximum value lies below 4.4\%, we replace the estimate
with the direct two-loop estimate of a 4.4\% error.

We next consider the error due to the finite volume (FV) of the lattice.
We estimate this from the difference between results on the C3 and C3-2
ensembles, which differ only in their spatial volumes.
This is not entirely satisfactory, since we do not use coarse lattices
in our final continuum-chiral extrapolation.
However, we stress that the dominant FV error, as estimated by SChPT,
comes from valence pions propagating to adjacent periodic volumes.
This is because the arguments of the chiral logarithms of Eq.~(\ref{eq:f0}) 
are the squared masses of valence pions, $X_B$.
Since on each ensemble we are extrapolating to the physical valence quark masses,
the dominant FV effects are present, even though on ensembles C3 and C3-2
we are far from the physical values of $L_P$ and $a$.
In our calculation of $B_K$, we have used the comparison of doing the X-fits with
finite- and infinite-volume SChPT forms as an alternative estimate of the FV 
error~\cite{Kim:2011qg}.
However, this method is not useful for the gold-plated combinations, since they
do not contain NLO chiral logarithms.

Our method of estimating systematic errors arising from the X fits 
has been described in Sec.~\ref{sec:fit-su2}.
We repeat the entire analysis using different priors for the X-fits,
and using the ES method. Each leads to a change in the final values of
the quantities of interest. We combine the fractional shifts in quadrature
to obtain our total systematic error.

Our method of estimating the systematic error arising from Y fits,
as noted above, is to repeat the entire analysis (including the
continuum-chiral extrapolation) using quadratic, as apposed to
linear, functional forms. 
This differs slightly from the estimate we used in Ref.~\cite{Bae:2014sja},
where we used the shift in the quantities on a 
specific MILC ensemble.
The Y fit errors turn out to be of comparable size to those from X fits,
ranging up to 2\%.

The remaining two systematic errors are very small, and have
essentially no impact on the total error. We include them for
completeness.  The first concerns the value of the pion decay constant
$f$ that we use in the chiral logarithms of Eq.~(\ref{eq:f0}). At NLO
we could equally well use the physical value $f_\pi =
130.41\MeV$~\cite{Beringer:1900zz} or the value in the chiral limit,
$f_\pi \approx 124.2\MeV$~\cite{Bazavov:2009bb}.  In practice we use
$f = 132\MeV$ (close to the physical value) for our central value, and
repeat the entire analysis using $f=124.2\MeV$ (the chiral-limit
value) to estimate the systematic error.
In fact, only $B_K$ is sensitive to this choice, since the gold-plated
combinations contain no NLO chiral logarithms.
Thus the 0.1\% error that results in $B_K$ propagates unchanged into
all of the BSM $B$-parameters.

Finally, the parameter we use to set the scale, $r_1$, has an error which
propagates into all the final results.
To estimate this, we repeat the entire analysis with
the central value for $r_1$ replaced by $r_1 \pm \sigma_{r_1}$, 
and quote the maximum difference in each quantity as the systematic error.
The resulting errors are very small ($\sim 0.1-0.35\%$), reflecting
the fact that the $B$-parameters are dimensionless.

\section{Comparisons and Outlook \label{sec:conclude}}

%
\begin{table*}[tbhp]
  \caption{Comparison of the BSM $B$-parameters at renormalization scale $\mu=3\GeV$ obtained using different fermion discretizations.
  The RBC-UKQCD results using domain-wall fermions are
    RBC-UK (2012)~\cite{Boyle:2012qb} and the preliminary results 
    (with incomplete error budget) of
    RBC-UK (2015)~\cite{Hudspith:2015wev}.
    %
    The ETM collaboration results using twisted-mass fermions are
    ETM (2012)~\cite{Bertone:2012cu} and  
    ETM (2015)~\cite{Carrasco:2015pra}.
    N.A. means ``not available''.
  \label{tab:comp}}
\begin{ruledtabular}
\begin{tabular}{ l | l l l l l }
      & SWME (this work) & RBC-UK (2012) & RBC-UK (2015) &  ETM (2012) & ETM (2015) \\
\hline
$B_K$              & \ben{0.519(4)(26)} & 0.53(2) & \wlee{0.53(1)} & 0.51(2) & 0.51(2)  \\
$B_2$              & 0.525(1)(23) & 0.43(5) & \wlee{0.49(2)} & 0.47(2) & 0.46(3)  \\
$B_3^\text{SUSY}$  & \ben{0.773(6)(35)} & 0.75(9) & \wlee{0.74(7)} & 0.78(4) & 0.79(5)  \\
$B_3^\text{Buras}$ & \ben{0.360(4)(16)} & N.A.    & N.A.    & N.A.    & N.A.     \\
$B_4$              & \ben{0.981(3)(62)} & 0.69(7) & \wlee{0.92(2)} & 0.75(3) & 0.78(5)  \\
$B_5$              & \ben{0.751(7)(68)} & 0.47(6) & \wlee{0.71(4)} & 0.60(3) & 0.49(4)  \\
\end{tabular}
\end{ruledtabular}
\end{table*}

\begin{figure*}[tbhp]
  \renewcommand{\subfigcapskip}{-0.55em}
  \subfigure[$B_4(3\GeV)$]{
    \label{fig:B4:3GeV}
    \includegraphics[width=0.48\textwidth]
                    {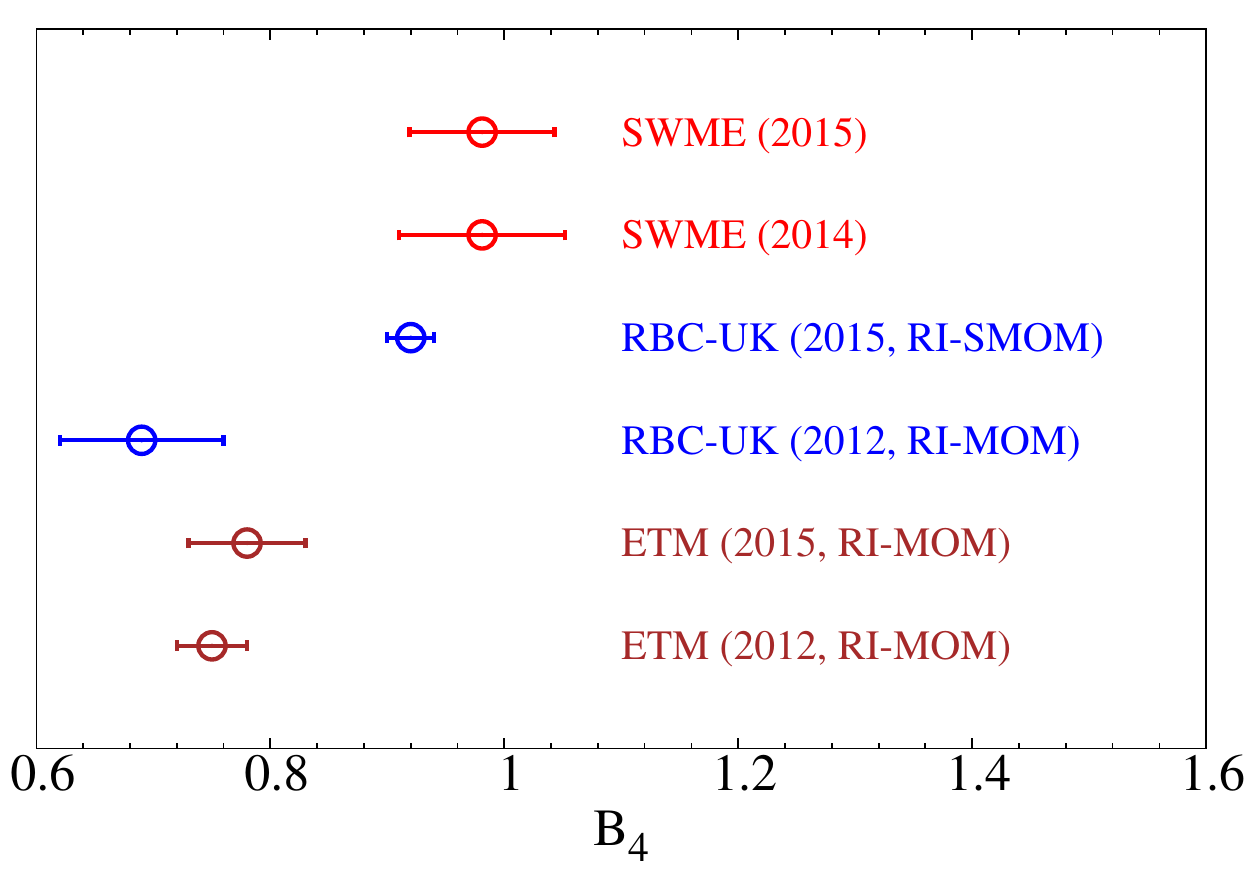}
  }
  \hfill
  \subfigure[$B_5(3\GeV)$]{
    \label{fig:B5:3GeV}
    \includegraphics[width=0.48\textwidth]
                    {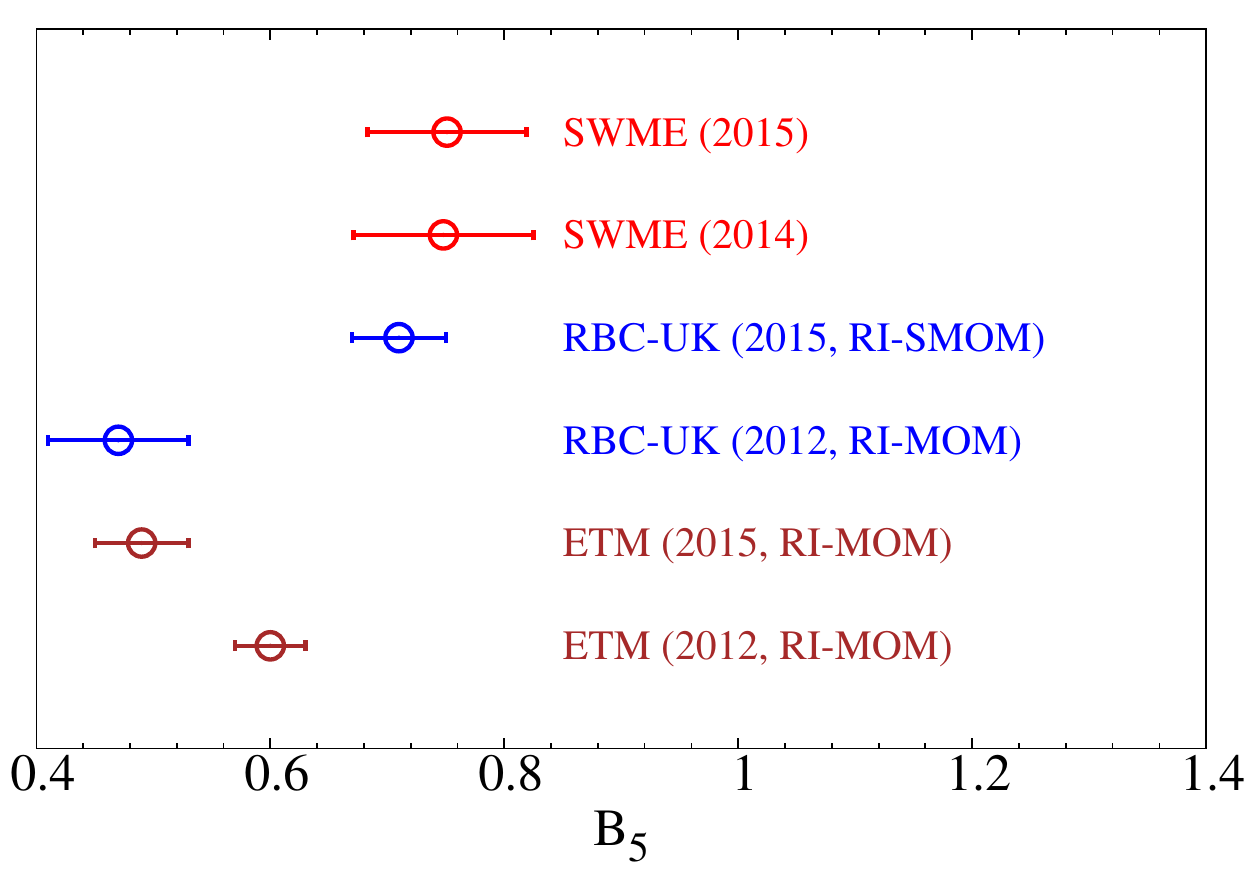}
  }
  \caption{ Comparison of results for $B_4$ and $B_5$ at $\mu=3\GeV$.
\sharpe{The references for the points are, proceeding from top to bottom,
this work (SWME 2015), \cite{Jang:2014aea} (SWME 2014),
\cite{Hudspith:2015wev} (RBC-UK 2015),
\cite{Boyle:2012qb} (RBC-UK 2012),
\cite{Carrasco:2015pra} (ETM 2015)
and \cite{Bertone:2012cu} (ETM 2012).}
}
\label{fig:cmp:B4+B5}
\end{figure*}

In Table~\ref{tab:comp} \sharpe{and Fig.~\ref{fig:cmp:B4+B5}}
we compare our results for the B-parameters
to those from other collaborations. This is done at $\mu=3\GeV$ since results
from all collaborations are available at this choice of renormalization scale.
The RBC-UKQCD collaboration uses $N_f=2+1$ light flavors of
domain wall quarks, and NPR
for the matching between lattice and continuum theories.
In 2012, RBC-UKQCD used the RI-MOM scheme for this matching~\cite{Boyle:2012qb},
while the preliminary 2015 results are obtained using several RI-SMOM schemes, 
in the spirit of Ref.~\cite{Sturm:2009kb}.
Both schemes are connected to the $\MSbar$ scheme using one-loop perturbation theory.
The ETM collaboration uses twisted-mass Wilson quarks.
The original results from 2012 were with $N_f = 2$ light sea quarks 
and a quenched valence strange quark~\cite{Bertone:2012cu},
while the 2015 results are from an $N_f=2+1+1$ simulation including
both strange and charmed sea quarks~\cite{Carrasco:2015pra}.
Both ETM calculations match lattice and continuum operators using NPR in the RI-MOM scheme.

Both RBC-UKQCD and ETM results are quoted using the so-called SUSY basis
of BSM four-fermion operators~\cite{Gabbiani:1996hi}.
The only BSM $B$-parameter which differs from that in the
basis of Buras {\em et al.} (Ref.~\cite{Buras:2000if}) that we use is $B_3$,
\begin{align}
\label{eq:susy-buras}
\begin{split}
B_3^\text{SUSY} &= -\frac{3}{2}B_3^\text{Buras} + \frac{5}{2}B_2^\text{Buras}
\,.
\end{split}
\end{align}
We use this equation to determine our result for $B_3^\text{SUSY}$ 
quoted in Table~\ref{tab:comp}.

For completeness, we note that our 2013 results for the BSM $B$-parameters
(Ref.~\cite{Bae:2013tca}) are superseded and corrected by our present results.\footnote{%
This does not apply to $B_K$, for which our present result is essentially the same as
that from Ref.~\cite{Bae:2013tca}.}
We now have significantly more ensembles, allowing a better controlled
continuum-chiral extrapolation. This addition required us to change
from $\tilde{F}_1$ fits to $\tilde{F}_4$ fits for $G_{23}$ and
$G_{45}$, which, as shown above, significantly changes the central
values for these quantities.  In addition, we found an error in our RG
running due to the use of an incorrect two-loop contribution to the
pseudoscalar anomalous dimension [needed for the denominators of the
  BSM $B$-parameters---see Eq.~(\ref{eq:def-B_i})].  Correcting this
error leads to $\sim 5\%$ reductions in all the BSM $B$-parameters.  A
detailed description of the effect of these changes is given in
Ref.~\cite{Jang:2014aea}.  The overall effect is that our new results
for $B_2$, $B_3^\text{SUSY}$, $B_4$ and $B_5$ are reduced by about
5\%, 3\%, 5\% and 12\%, respectively, compared to those in
Ref.~\cite{Bae:2013tca}.

Table~\ref{tab:comp} shows that the results for $B_K$, $B_2$ and
$B_3^\text{SUSY}$ are consistent across all calculations, with all
results having comparable errors.  By contrast, there are significant
differences for $B_4$ and $B_5$, \sharpe{as one can see most clearly from
  Fig.~\ref{fig:cmp:B4+B5}}.  The preliminary results from RBC-UKQCD
(2015) using the intermediate RI-SMOM schemes are consistent with our
results, while those using the intermediate RI-MOM scheme [RBC-UK
  (2012), ETM (2012) and ETM (2015)] differ significantly.  For
example, the ETM (2015) results for $B_4$ and $B_5$ differ from our
results by $2.6\sigma$ and $3.2\sigma$, respectively.

The pattern of results in the Table suggests that the ultimate source
of these differences may well be the matching from lattice matrix
elements to those in the continuum $\MSbar$ scheme. 
In our calculation, this error is due to the truncation of matching
factors at one-loop order.
For $B_4$ and $B_5$ (the two $B$-parameters which
differ from the results obtained using the RI-MOM scheme)
our error estimate is taken as
the difference between fits using $\tilde F_1$ and $\tilde F_4$ fit forms
(see Figs.~\ref{fig:G24:F1+F4:2GeV} and \ref{fig:G45:F1+F4:2GeV}).
While we consider this to be a conservative estimate,
we cannot rule out that it is an underestimate
due to unexpectedly large $\alpha^2$ terms in the matching factors.
In the case of the calculations using the NPR method, 
the significant differences between results obtained using 
RI-MOM and RI-SMOM schemes indicate an
underestimate of the associated systematic errors.
This could be a problem specifically related to the RI-MOM scheme,
where one must subtract unwanted contributions from pion poles,
a source of systematic errors absent in the RI-SMOM schemes~\cite{Lytle:2013oqa}.
Or it could be due to large truncation errors in the relation
between one or both of these schemes and the $\MSbar$ scheme.

Clearly these issues require further investigation.
One possibility is for 
all the calculations to use the same intermediate scheme such as RI-SMOM
and then to directly compare results in that scheme.
This reduces the dependence on perturbation theory as one does not
need to to match to the $\MSbar$ scheme.
One would still need to evolve between different scales in the RI-SMOM
scheme, but this could also, ultimately, 
be done non-perturbatively~\cite{Arthur:2010ht}.
To these ends we are pursuing the implementation of NPR using staggered
fermions \cite{Lytle:2013qoa, Kim:2013bta, Jeong:2014jia}.

\begin{acknowledgments}
%
We thank Peter Boyle, Nicolas Garron, Jamie Hudspith and Andrew Lytle  for
discussions of the RBC-UKQCD results and comments on the manuscript.
We would also like to express our sincere gratitude to Claude Bernard and 
MILC collaboration for private communications.
C.~Jung is supported by the US DOE under contract DE-AC02-98CH10886.
Jangho Kim is supported by Young Scientists Fellowship through
National Research Council of Science \& Technology (NST) of KOREA.
The research of W.~Lee is supported by the Creative Research
Initiatives Program (No.~2015001776) of the NRF grant funded by the
Korean government (MEST).
W.~Lee would like to acknowledge the support from the KISTI
supercomputing center through the strategic support program for the
supercomputing application research (No.~KSC-2014-G3-003).
The work of S.~Sharpe is supported in part by the US DOE grants
no.~DE-FG02-96ER40956 and DE-SC0011637.
Part of computations were carried out on the DAVID GPU clusters at
Seoul National University.
\end{acknowledgments}

\bibliography{ref} 

\end{document}